\documentclass[aps,prl,longbibliography,lengthcheck,superscriptaddress,nopreprintnumbers,reprint,groupedaddress,showkeys]{revtex4-1}
\renewcommand{\tablename}{Table}
\renewcommand{\thetable}{\arabic{table}}
\renewcommand{\figurename}{Fig.}
\renewcommand{\thefigure}{\arabic{figure}}
\renewcommand{\theequation}{\arabic{equation}}
\usepackage{graphicx}
\usepackage[percent]{overpic}
\usepackage{subcaption}
\usepackage{ragged2e}
\usepackage{dcolumn}
\usepackage{bm}        
\usepackage{amssymb}   

\usepackage{bm}
\usepackage[usenames ,dvipsnames]{xcolor}
\usepackage{color}
\usepackage{url}
\usepackage{dsfont}
\usepackage{slashed}
\usepackage{epsfig}
\usepackage{multirow,array}
\usepackage{float}
\usepackage{fancyhdr}
\usepackage{indentfirst}
\usepackage{cancel}
\usepackage{tabularx}
\usepackage{simplewick}
\usepackage{amsmath}
\usepackage[normalem]{ulem}
\usepackage{slashed}
\usepackage{upgreek}
\usepackage{float}
\usepackage{xtab,afterpage}
\RequirePackage[colorlinks=true
  ,urlcolor=Green
  ,anchorcolor=blue
  ,citecolor=orange
  ,filecolor=blue
  ,linkcolor=blue
  ,menucolor=blue
  ,pagecolor=blue
  ,linktocpage=true
  ,pdfproducer=medialab
  ,pdfa=true
]{hyperref}

\newcommand{\jcap}{JCAP}

\begin{document}

\title{Self-interacting dark matter with mass segregation: a unified explanation of dwarf cores and small-scale lenses}

\author{Daneng Yang }
\email{yangdn@pmo.ac.cn}
\author{Yi-Zhong Fan }
\email{yzfan@pmo.ac.cn}
\author{Siyuan Hou }
\author{Yue-Lin Sming Tsai }
\email{smingtsai@pmo.ac.cn}

\affiliation{$^1$Purple Mountain Observatory, Chinese Academy of Sciences, Nanjing 210033, China}
\affiliation{$^2$School of Astronomy and Space Sciences, University of Science and Technology of China, Hefei 230026, China}

\begin{abstract}
In two-component self-interacting dark matter (SIDM) models with inter-species interactions, mass segregation arises naturally from collisional relaxation, enhancing central densities and gravothermal evolution. We demonstrate that models with velocity-dependent interactions, both within and between species, can connect several small-scale observations while remaining consistent with cluster-scale constraints. This combination enables core formation in dwarf halos, where the presence of baryons increases the inner densities and enhances the predicted strong lensing signatures. Using cosmological and controlled simulations alongside an accurate parametric model, we present proof-of-principle examples showing that this framework can explain the structure of dark perturbers observed in strong lensing systems, and can enhance the efficiency of small-scale lenses by a factor of a few, in line with the excess reported in galaxy–galaxy strong-lensing observations. Importantly, mass segregation can enhance the Einstein radii of SIDM halos relative to their cold dark matter (CDM) counterparts, overcoming a key challenge in one-component SIDM scenarios. Our results present mass segregation in two-component SIDM as a self-consistent, testable framework with the potential to address multiple small-scale challenges in structure formation.
\end{abstract}

\keywords{Self-interacting dark matter, Mass segregation, Strong lensing, N-body simulation, Parametric model}

\maketitle

\section{1. I\MakeLowercase{ntroduction}}

Observations of large-scale structure have established the standard cosmological model with cold dark matter (CDM).
However, on small scales, persistent discrepancies continue to challenge this framework, motivating alternative interpretations~\cite{Bullock:2017xww,Salucci:2018hqu,Gentile:2004tb,vanDokkum:2022zdd,Bonaca:2018fek}.
Among these, self-interacting dark matter (SIDM) has long been considered a promising candidate \cite{Spergel:1999mh,Tulin:2017ara}.
Through gravothermal evolution, SIDM can naturally produce both cored and cuspy halo density profiles, providing a flexible mechanism to explain the diversity of galactic rotation curves, the extreme central densities observed in ultra-diffuse and ultra-compact galaxies, and the early formation of supermassive black holes \cite{Kaplinghat:2015aga,Robertson:2018anx,despali:2018zpw,essig:2018pzq,santos-santos191109116,turner:2020vlf,Fischer:2020uxh,slone:2021nqd,Zeng:2021ldo,Gilman:2021sdr,correa:2022dey,yang220503392,Adhikari:2022sbh,outmezguine:2022bhq,silverman:2022bhs,yang:2023jwn,zhong:2023yzk,Zeng:2023fnj,Zhang:2024qem,Zhang:2024ggu,Ragagnin:2024deh,Nadler:2025jwh,Kong:2025irr,Jiang:2025jtr,Kamionkowski:2025uae,Kamionkowski:2025fat}.
It has also been proposed as a solution to the final-parsec problem in binary black hole mergers \cite{Alonso-Alvarez:2024gdz}.
Recently, 
Zhang et al. 
\cite{Zhang:2025bju} revealed an unexpected clustering pattern in dwarf galaxies, favoring a tentative self-interaction cross section of $\sigma/m \sim 0.3~\rm cm^2/g$ in their host halos. 

Despite accumulating evidence, a unified SIDM explanation for all small-scale phenomena remains an open challenge. 
Observations of dwarf galaxy clustering suggest that a population of these galaxies has growing cores, while the reconstruction of compact substructures inferred from strong lensing images, such as those observed in the SDSS J0946+1006 and JVAS B1938+666 systems~\cite{2010MNRAS.408.1969V,2012Natur.481..341V,Despali:2024ihn,Ballard:2023fgi,Enzi:2024ygw,Powell:2025rmj}, indicates the presence of core-collapsed dwarf halos.
Moreover, core collapse alone cannot fully account for the elevated abundance of small-scale lenses reported in galaxy-galaxy strong lensing (GGSL) by Meneghetti et al~\cite{Meneghetti:2020yif,Meneghetti:2022apr,Meneghetti:2023fug}, which exceeds CDM expectations by a factor of 3 to 6. 
Ref.~\cite{yang:2021kdf} further finds that SIDM halos with baryons primarily enhance two-image lensing events, rather than the four-image ones observed.

In this work, we demonstrate that SIDM with mass segregation provides a natural unified framework for resolving these tensions. 
Mass segregation in multi-component SIDM models arises as a consequence of collisions among particles of different masses. 
In the two-component case, inter-species scatterings lead to energy transfer from the heavier to the lighter species, resulting in denser inner halos~\cite{Yang:2025dgl,Patil:2025nmj}. 
This process is negligible in CDM due to inefficient gravitational relaxation among particles (unlike stars~\cite{Zhong:2025epi}), but becomes significant in SIDM, where collisional interactions dominate.
Previous simulations at Milky Way scales~\cite{Yang:2025dgl} demonstrated that this mechanism reshapes inner halo profiles under strong scattering. 
Here, we extend this framework to simulations across dwarf to cluster scales, developing a conditioned parametric model to accurately capture gravothermal evolution near the resolution limit.
We show that even modest inter-species interactions can induce mass segregation, increasing central densities while producing cores in dwarf halos. 
Crucially, unlike in one-component SIDM, the mass enclosed within the Einstein radius can also grow, amplifying strong lensing signals.
Our results point to a unified scenario in which both cored dwarf galaxies and cuspy cluster substructures emerge naturally from SIDM with mass segregation.

\section{2. A \MakeLowercase{two-component} SIDM \MakeLowercase{model with mass segregation}}

\begin{figure}[htbp]
  \centering
  \includegraphics[width=7.2cm]{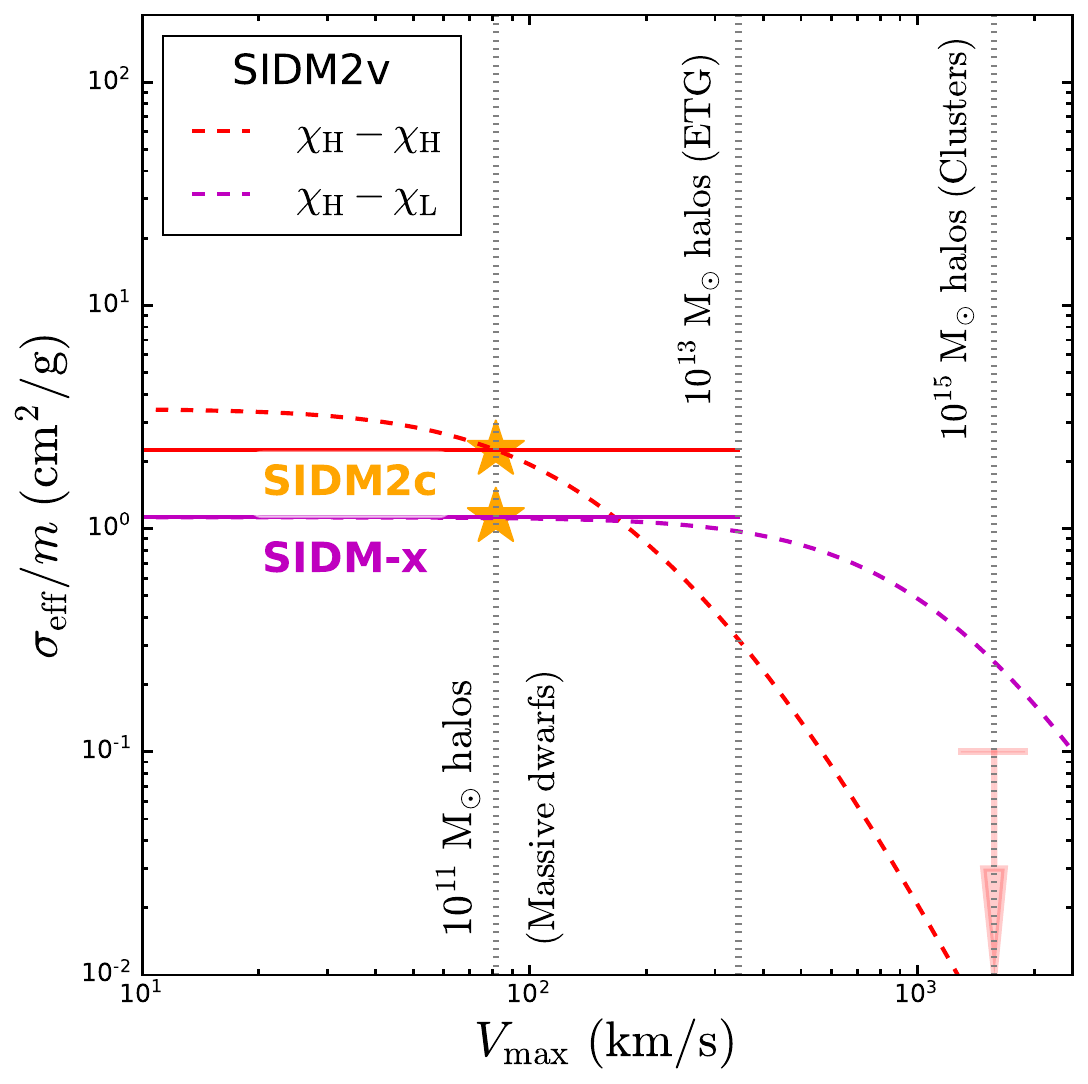}
  \caption{\label{fig:xs} Effective cross section per mass $\sigma_{\rm eff}/m$ as a function of $V_{\rm max}$ for various SIDM models. 
The self-interaction among the heavier component ($\chi_{\rm H}-\chi_{\rm H}$) and the inter-species interaction ($\chi_{\rm H}-\chi_{\rm L}$) in the \texttt{SIDM2v} model are represented by red and magenta solid curves. 
This model is featured by its capability of reproducing SIDM effects analogous to one dark matter component with $\sigma/m = 0.3 ~{\rm cm^2/g}$ (\texttt{SIDM03}) in massive dwarf halos of mass around $M_{\rm h} = 10^{11}~\rm M_{\odot}$, as favored by the findings of Ref.~\cite{Zhang:2025bju}.
The two orange stars feature the two kinds of strengths in the two-component \texttt{SIDM2c} model, which takes constant effective cross sections as in \texttt{SIDM2v} in a $M_{\rm h} = 10^{11}~\rm M_\odot$ halo with median concentration. 
For $V_{\rm max} \gtrsim 200~\mathrm{km/s}$, the inter-species interaction dominates over the intra-species interactions and remains significant up to the cluster scale at $V_{\rm max} \approx 1600~\mathrm{km/s}$. 
In clusters, the coexistence of both types of scattering yields density profiles consistent with observations. Particularly, the intra-species interaction remains well below $0.1~\rm cm^2/g$, as indicated by the light-red arrow.  
}
\end{figure}

Two-component dark matter models naturally emerge in many particle models, such as those that mirror the asymmetries of the baryonic sector, where protons and electrons are stable and interact with each other~\cite{Kaplan:2009ag,Bansal:2022qbi,Petraki:2011mv,Yang:2025dgl}.
To incorporate mass segregation, we consider a two-component SIDM model in which the heavier species, $\chi_{\rm H}$, has mass $m_{\rm H} = 3 m_{\rm L}$, with $\chi_{\rm L}$ denoting the lighter component. We assume equal number densities and identical initial spatial distributions for both species. 
We introduce a model, \texttt{SIDM2v}, featuring velocity-dependent interactions for both identical (intra-species) and distinguishable (inter-species) scatterings. While both cross sections decrease with velocity, the inter-species interaction dominates at high velocities ($V_{\rm max} \gtrsim 200~\mathrm{km/s}$). 
Specifically, we adopt the Møller and Rutherford parametrizations for identical and distinguishable particle scatterings, following equations in Refs.~\cite{Yang:2025dgl,yang220503392}. The intra-species interaction is characterized by $\sigma_0/m_{\rm H} = 6.89~\rm cm^2/g$ and a velocity scale $w = 275~\rm km/s$ (denoted $w_{\rm H}$), while the inter-species interaction uses $\sigma_0/m_{\rm H} = 1.125~\rm cm^2/g$ and $w = 2200~\rm km/s$. We assume that the same particle mediates all intra-species interactions, so that the $\chi_{\rm L}-\chi_{\rm L}$ interaction shares the same $\sigma_0$, with a velocity scale $w=(m_{\rm H}/m_{\rm L})w_{\rm H}$. 

Fig.~\ref{fig:xs} illustrates our model in terms of effective cross sections, defined as velocity- and angle-averaged differential cross sections weighted by a $v^5\sin^2\theta$ kernel and convolved with the dark matter velocity distribution. The $x$-axis corresponds to halo velocity scales via their maximum circular velocity, $V_{\rm max}$. 
The two orange stars indicate the intra- and inter-species interactions in a \texttt{SIDM2c} model without velocity dependence. Their strengths are set by matching the effective cross sections of \texttt{SIDM2v} in a benchmark halo of mass $10^{11}~\rm M_{\odot}$, yielding $\sigma_{0}/m = 2.25~\rm cm^2/g$ for intra-species and $\sigma_{0}/m = 1.125~\rm cm^2/g$ for inter-species interactions.
This setup leads to time-dependent core formation in most dwarf halos of mass $\sim 10^{11}~\rm M_{\odot}$, comparable to one-component $\sigma/m = 0.3~\rm cm^2/g$, as motivated by Ref.~\cite{Zhang:2025bju}. 

To examine small-scale effects in detail, we perform high-resolution simulations of isolated halos in \texttt{SIDM2c}, which replicate the relevant interaction regime in \texttt{SIDM2v}. 
We also consider a model, \texttt{SIDMx}, that includes only inter-species interactions from \texttt{SIDM2c}, representing an extreme scenario of mass segregation. Notably, both intra- and inter-species interactions are weak yet effective under \texttt{SIDM2v} in clusters. Inter-species interactions drive condensation, which competes with core formation induced by intra-species self-interactions, resulting in a density profile consistent with the $\sigma/m < 0.1~\rm cm^2/g$ constraint from cluster lensing observations~\cite{Andrade:2020lqq,Kaplinghat:2015aga,2019MNRAS.487.1905A}. We provide an additional supportive test in the Supplemental Material. 
Table~\ref{tab:model} summarizes the two-component SIDM models simulated in this work.

\begin{table}[h]
    \centering
    \caption{\label{tab:model} Benchmark models of two-component SIDM with mass segregation.}
    \begin{tabular}{cccc}
        \hline
        Model & Scatterings & $\frac{\sigma_0}{m}$ relation  & $w$ relation  \\
        \hline
        \multirow{3}{*}{SIDM2c} & $\chi_{\rm H}{\texttt-} \chi_{\rm H}$  & $\frac{\sigma_{\rm H}}{m_{\rm H}}=4.5\rm \frac{cm^2}{g}$ & --  \\
        & $\chi_{\rm L}{\texttt-} \chi_{\rm L}$  & $\frac{\sigma_{\rm L}}{m_{\rm L}}=\frac{1}{3}\frac{\sigma_{\rm H}}{m_{\rm H}}$ & -- \\
         & $\chi_{\rm H}{\texttt-} \chi_{\rm L}$  & $\frac{\sigma_{\rm x}}{m_{\rm H}}=\frac{1}{4}\frac{\sigma_{\rm H}}{m_{\rm H}}$ & -- \\ 
        \hline
        \multirow{3}{*}{SIDM2v} & $\chi_{\rm H}{\texttt-} \chi_{\rm H}$  & $\frac{\sigma_{\rm H}}{m_{\rm H}}=6.89\rm \frac{cm^2}{g}$ & $w_1=275\ \rm\frac{km}{s}$  \\
         & $\chi_{\rm L}{\texttt-} \chi_{\rm L}$  & $\frac{\sigma_{\rm L}}{m_{\rm L}}=\frac{1}{3}\frac{\sigma_{\rm H}}{m_{\rm H}}$ & $w_2 = 3 w_1$ \\
         & $\chi_{\rm H}{\texttt-} \chi_{\rm L}$  & $\frac{\sigma_{\rm x}}{m_{\rm H}}=1.125\rm \frac{cm^2}{g}$ & 2200 $\rm \frac{km}{s}$ \\ 
        \hline
        SIDMx & $\chi_{\rm H}{\texttt-} \chi_{\rm L}$  & $\frac{\sigma_{\rm x}}{m_{\rm H}}=1.125\rm \frac{cm^2}{g}$ & -- \\ 
        \hline
    \end{tabular}
\vspace{0.5em}
\parbox{\linewidth}{ 
\footnotesize
The $\sigma_0/m$ and $w$ parameters are introduced to parametrize the scattering cross sections, following Ref.~\cite{Yang:2025dgl,yang220503392} and considering Møller (Rutherford) scatterings for identical (distinguishable) particles. For identical particle scattering, a phase space suppression factor of $1/2$ is included to enable comparison with distinguishable inter-species scatterings, yielding the $\sigma_0/m=4.5/2{\rm cm^2/g}=2.25~\rm cm^2/g$ in the \texttt{SIDM2c} case.
}
\end{table}

\section{3. N\MakeLowercase{umerical methods and modeling}}

To accurately model signatures including baryonic and environmental effects, we employ controlled and cosmological simulations, combined with a parametric model for SIDM2c. The latter is constructed to obtain accurate halo profiles beyond the simulation resolution limit.
We use the parametric model to mitigate resolution limitations in mapping CDM merger histories to SIDM2c profiles, and we adopt a conservative approach to the resulting lensing inferences.

Our simulations are based on the SIDM module developed on top of the \texttt{Gadget-2} program~\cite{Springel0505010,Springel:2000yr} in Refs.~\cite{yang220503392,yang:2022mxl}, extended to handle two dark matter species with unequal masses. This module has been validated in both isolated and cosmological contexts. 
To establish two-component initial conditions without altering the system's dynamical properties, we randomly select half of the simulation particles to be 50\% more massive, and the other half to be 50\% less massive, preserving the total system mass.
For isolated halos, we generate Navarro-Frenk-White (NFW) profiles~\cite{1997apj...490..493n} in hydrostatic equilibrium using \texttt{SpherIC}~\cite{GarrisonKimmel:2013aq}, simulating with $N=10^6$ particles per run. The softening length is chosen to be a few $R_{\rm vir}/\sqrt{N}$~\cite{Power:2002sw}, where $R_{\rm vir}$ is the virial radius. 
A stellar component, modeled by the Hernquist profile~\cite{1990ApJ...356..359H}, is included in the study with baryons. They are simulated as collisionless particles with the same mass as the dark matter particles. In the Supplemental Material, we provide a summary of the controlled simulations in a table. 
For cosmological simulations, we initialize a periodic box of size $400~{\rm Mpc}/h$, and perform zoom-in simulations of a $20~{\rm Mpc}/h$ region centered on a host halo of mass $10^{15}~{\rm M_{\odot}}/h$, with a softening length of one comoving ${\rm kpc}/h$, where $h$ is the dimensionless Hubble parameter. The highest-resolution particles have average mass $m_0 \equiv (m_1 + m_2)/2 \approx 7\times 10^7~{\rm M_{\odot}}/h$.
As shown in Ref.~\cite{Yang:2025dgl}, mass segregation is negligible in CDM when particle masses are realistic, so we simulate CDM using a single-component model. 
We also apply a modified version of the \texttt{Rockstar} halo finder~\cite{Behroozi:2011ju} to accommodate two-component dark matter, a method validated and applied in Ref.~\cite{Yang:2025dgl}. Using the resulting halo catalogs, we then build halo merger trees with the \texttt{consistent-trees} program~\cite{Behroozi11104370}.
For one-component SIDM in isolated halos, we employ the parametric framework of Ref.~\cite{yang:2023jwn}, which has been extensively tested across controlled and cosmological settings~\cite{Yang:2024uqb}.

To construct a parametric model for SIDM with mass segregation, we first show that isolated halos in our setup exhibit \emph{conditioned universality}. 
Within the conduction-fluid framework, the coupled transfer equations for the heavy (H) and light (L) dark matter species are
\begin{eqnarray}
\rho_{\rm H} \nu_{\rm H}^2 \left(\frac{D}{D t} \right) \ln\frac{\nu_{\rm H}^3}{\rho_{\rm H}} &=& -\frac{1}{4\pi r^2}\frac{\partial L_{\rm H}}{\partial r} - R, \\ \nonumber
\rho_{\rm L} \nu_{\rm L}^2 \left(\frac{D}{D t} \right) \ln\frac{\nu_{\rm L}^3}{\rho_{\rm L}} &=& -\frac{1}{4\pi r^2}\frac{\partial L_{\rm L}}{\partial r} + R, \\ \nonumber
\end{eqnarray}
where $\rho,\nu$ and $L$ denote density, velocity dispersion, and luminosity, respectively. 
Here, $D/Dt$ is the material derivative and $R$ the inter-species energy transfer rate. 

\begin{figure}[htbp]
  \centering
  \includegraphics[height=7.2cm]{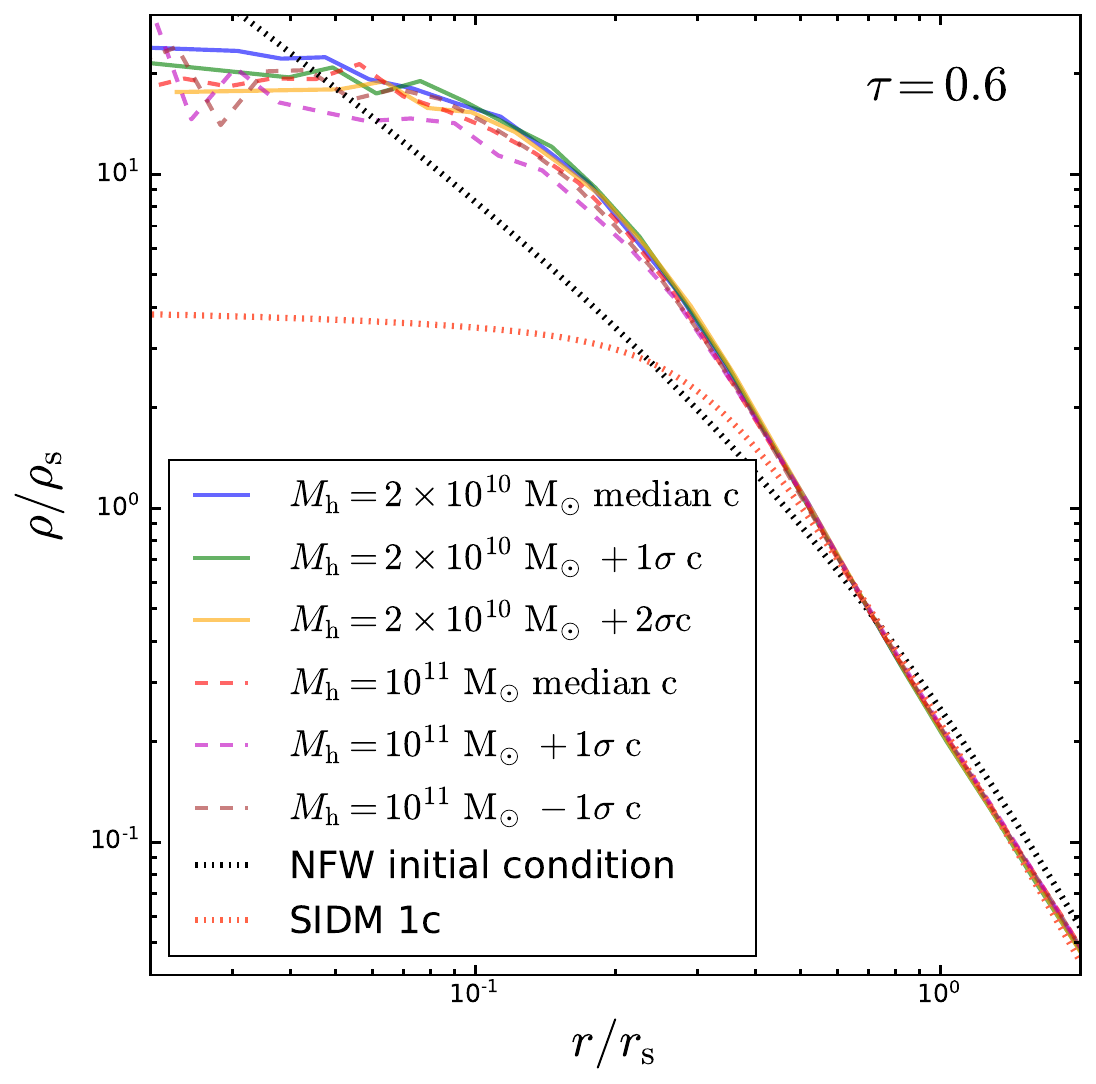}
  \caption{\label{fig:parametricM} Rescaled halo density profiles at 60\% of the core collapse time $\tau\equiv t/t_{\rm c}=0.6$, compared to a single-component (SIDM1c) model at the same evolution stage. The profiles of three halos of mass $2\times 10^{10}~\rm M_{\odot}$ (solid) and three of mass $10^{11}~\rm M_{\odot}$ (dashed) with median, $+1\sigma$, and $-1\sigma$ concentrations have almost identical shapes, illustrating the high quality of universality supporting the construction of the parametric model. 
}
\end{figure}

\begin{figure*}[htbp]
  \centering

  \begin{subfigure}{0.48\textwidth}
    \centering
    \begin{overpic}[height=7.2cm]{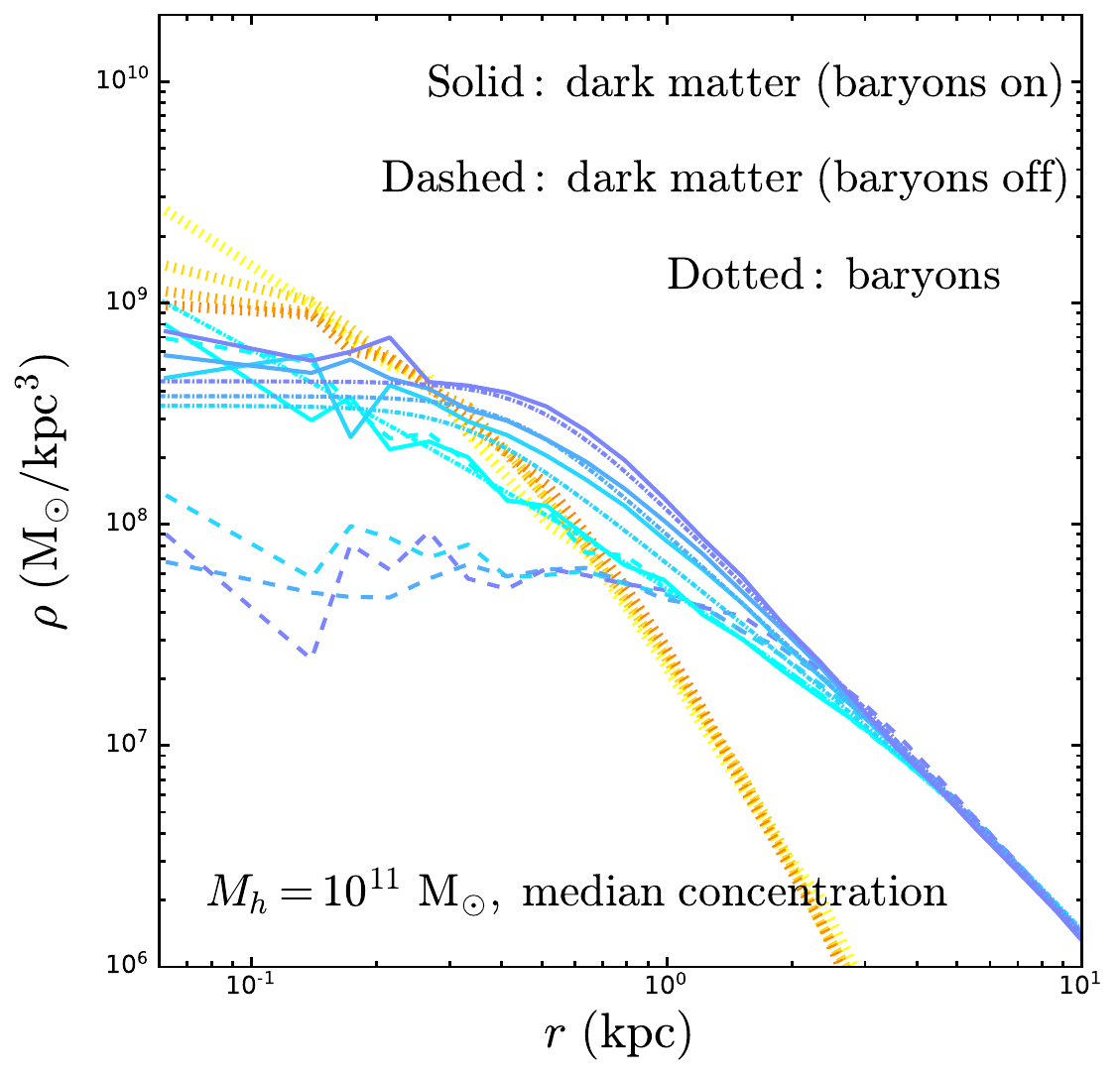}
      \put(3,92){\textsf{\textbf{(a)}}}
    \end{overpic}
  \end{subfigure}
  \hfill
  \begin{subfigure}{0.48\textwidth}
    \centering
    \begin{overpic}[height=7.2cm]{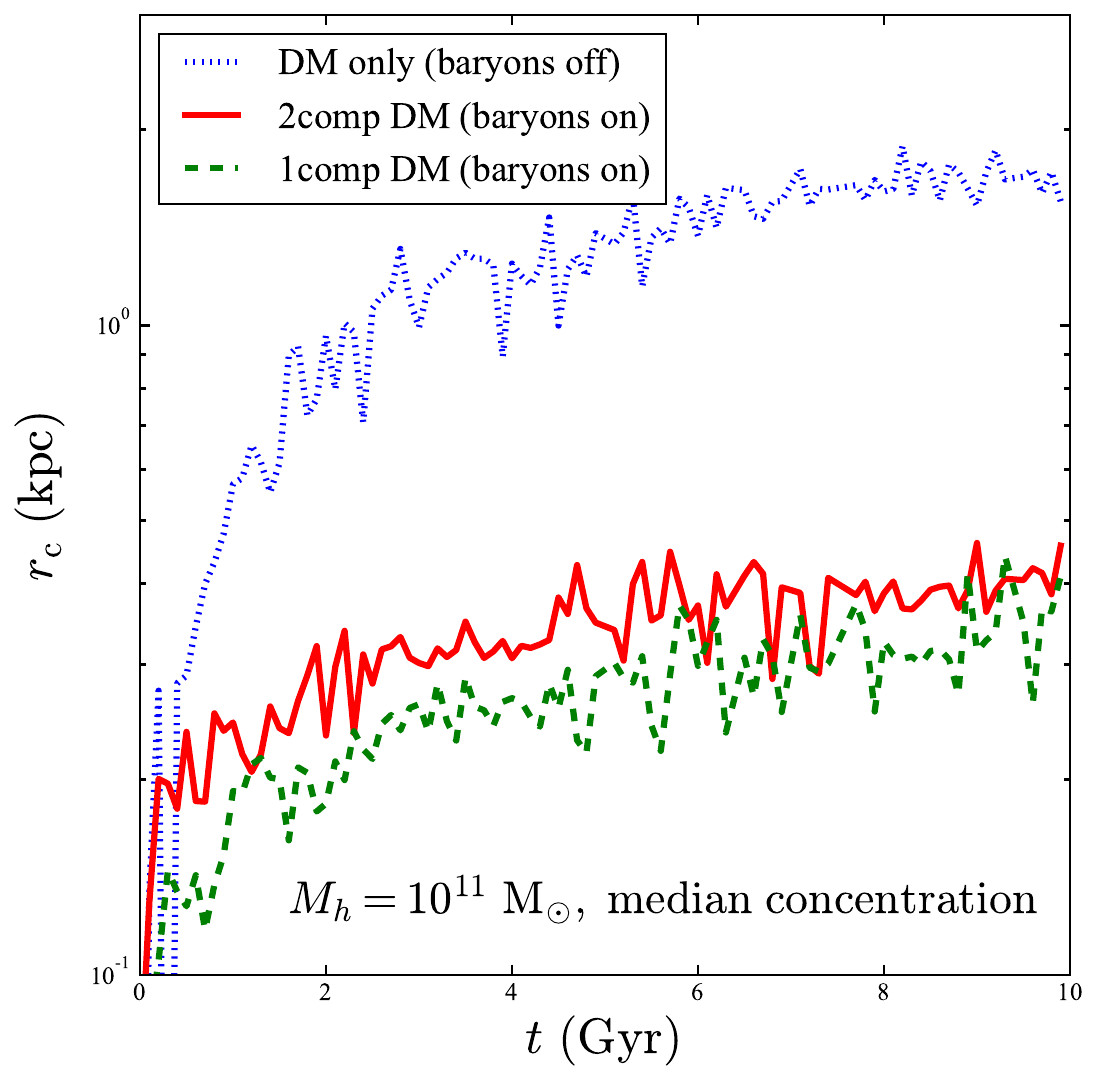}
      \put(3,92){\textsf{\textbf{(b)}}}
    \end{overpic}
  \end{subfigure}

  \caption{\label{fig:dwarfs} Dwarf halo evolution and core formation in one- and two-component SIDM. 
  (a) Density profiles for the median halo with a collisionless stellar component at $t=0$, $3$, $6$, and $9~$Gyr, with deeper colors corresponding to later times. Baryons (dotted tomato lines) induce higher DM densities (solid) compared to DM-only cases (dashed). Unlike DM-only evolution, central density increases even as the cores grow in size. The parametric model predictions are shown in dash-dotted lines. (b) Evolution of core size ($r_c$) for the median halo with (solid and dashed) and without (dotted) baryons. The core growth in \texttt{SIDM2c} remains comparable to that in \texttt{SIDM03} in the presence of baryons.
}
\end{figure*}

In the long-mean-free-path regime, the SIDM effect scales with the scattering rate, hence $L_{\rm H}, L_{\rm L}$ and $R$ are all proportional to the scattering cross section. 
In two-component models, the normalization of these quantities, parameterized by $\sigma_s/m$, can be absorbed into the evolution time, defining a rescaled evolution variable $t\sigma_s/m$. It follows that SIDM halos exhibit a universal evolutionary track when the mass ratio and the relative inter- to intra-species cross sections are fixed.

Fig.~\ref{fig:parametricM} illustrates the accuracy of this conditioned universality by showing that the rescaled halo density profiles have almost the same shapes if evolved under \texttt{SIDM2c} for 60\% of their core collapse times. We consider two halos of masses $2\times 10^{10}~\rm M_{\odot}$ and $10^{11}~\rm M_{\odot}$, with varied concentrations computed following the $c_{200}-M$ relation from Ref.~\cite{Dutton:2014xda} and assuming a logarithmic scatter of $0.11$ dex.
Compared with the density profile of a halo in one-component SIDM (SIDM 1c) at the same gravothermal phase ($\tau\equiv t/t_{\rm c}=0.6$, shown in dotted, tomato-colored curve), the profiles in \texttt{SIDM2c} have cored profiles and significantly higher inner densities. 

By fitting the universal density profile evolution, we construct an accurate parametric model for halos under \texttt{SIDM2c}. 
We also extend the model to capture leading baryon effect on the profile evolution, which is achieved by simply rescaling the core size in the DM-only case by a form factor introduced in Ref.~\cite{Yang:2024tba}. 
In Ref.~\cite{Meneghetti:2022apr,2021MNRAS.505.1458B,2021MNRAS.504L...7R,2024ApJ...970..143T}, it has been demonstrated that baryons have a significant effect on cluster subhalos. While simplified, our method captures baryons' contraction on \texttt{SIDM2c} evolution, the effect of stellar-to-halo mass ratio, and reduces to the DM-only case without baryons. 
We provide supplemental material to detail the model and demonstrate its robustness.

\begin{figure}[htbp]
  \centering
  \includegraphics[width=7.5cm]{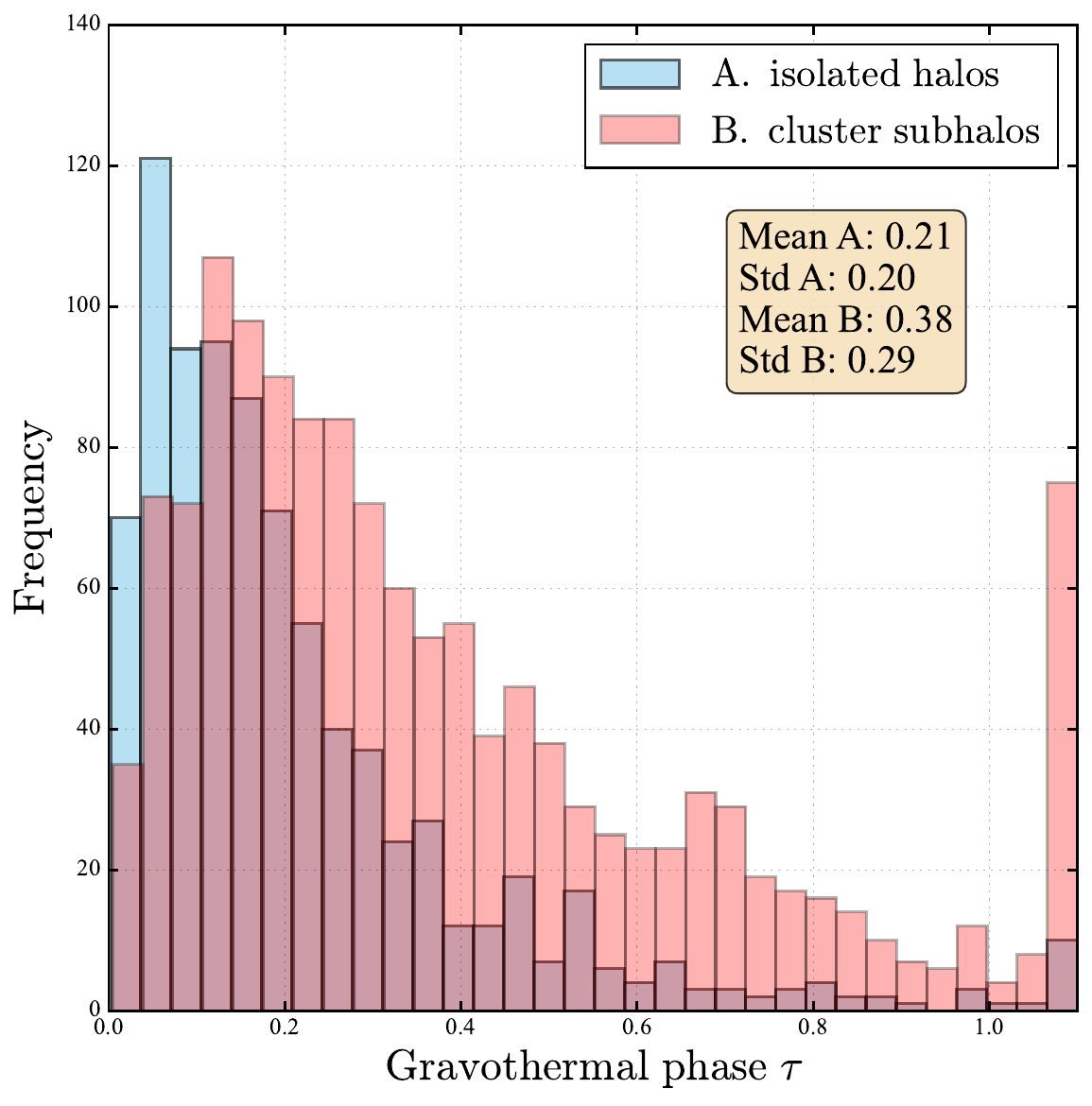}
  \caption{\label{fig:tau} Distribution of the gravothermal phase $\tau$ from parametric model predictions for the \texttt{SIDM2c} model. The sample includes halos of masses higher than $10^{10}~\rm M_{\odot}$ and within $5~\mathrm{Mpc}/h$ of the host cluster, separated into cluster subhalos (group~A, light red) within the cluster's virial radius and isolated halos (group~B, light blue). Means and standard deviations for both groups are shown in the inset. Most halos in both groups have cored inner profiles ($\tau < 0.6$). Cluster subhalos, however, show a greater frequency of core collapsed systems ($\tau > 1$), as a consequence of tidal stripping.
}
\end{figure}

\section{4. D\MakeLowercase{warf scale cores}}

Ref.~\cite{Zhang:2025bju} finds that diffuse dwarf galaxies exhibit unexpectedly stronger large-scale clustering than denser ones. This can be naturally explained in SIDM, where halos in the core-forming phase develop larger cores if they evolve longer. Since halos in overdense regions form earlier, they are more likely to host diffuse galaxies, leading to enhanced clustering. 
To demonstrate that our proposed two-component models can reproduce this behavior, we simulate isolated halos with mass $M = 10^{11}~\rm M_{\odot}$, which host the dwarfs in Ref.~\cite{Zhang:2025bju}, using \texttt{SIDM2c}.
These simulations show that the halos develop time-dependent cores, similar to those in the one-component model with $\sigma/m=0.3~\rm cm^2/g$ (\texttt{SIDM03}), as favored in that work.

We simulate the median-concentration halo with and without a collisionless stellar component, which follows median stellar-to-halo mass and size-mass relations. 
Fig.~\ref{fig:dwarfs}a presents density profiles for the median halo with and without this stellar component. 
While the halo densities in the baryonic cases (solid) are systematically higher than the DM-only cases (dashed), we found two crucial differences. 
One is that the stellar component develops cores in time, a feature directly addresses the observations. 
The other is that halos also have growing cores along with the increasing inner densities. 
This feature is precisely the opposite of the dark matter only case, where the core formation is accompanied with a decreasing inner halo density. 

To track core growth clearly, we fit the halo density profiles with a $\beta4$ model, defined as $\rho_{\beta4} = \rho_{\rm s}/((r/r_{\rm s})^4+(r/r_c)^4)^{1/4}/(1+r/r_{\rm s})^2$~\cite{Gilman:2021sdr,yang:2023jwn}, and plot the resulting core size $r_c$ evolution in Fig.~\ref{fig:dwarfs}b.
After including baryons, the core size is significantly reduced due to gravitational contraction. 
Crucially, the core growth in \texttt{SIDM2c} appears close to that of the one-component \texttt{SIDM03} after incorporating baryons. 

In Fig.~\ref{fig:tau}, we present the distribution of the gravothermal phase $\tau$ from parametric model predictions to show the dominance of cored halos in the dwarf halo population.  
The sample is selected by considering halos with masses higher than $10^{10}~{\rm M_{\odot}}/h$ and within 5 Mpc/h of the cluster host halo in the cosmological zoom-in simulation.  
For dwarf clustering analysis, we consider isolated halos (group B, light blue) and therefore exclude cluster subhalos (i.e., those residing within the virial radius of the cluster halo).  
These halos are predominantly core forming, as their $\tau$ values are low, with mean and standard deviation both close to 0.2~\cite{yang:2023jwn}.  
The cluster subhalos (group A, light red) have higher $\tau$ values because their progenitor halos are more massive. The mean of the distribution is shifted to 0.38, and there is a small population of halos that have undergone deep core collapse ($\tau>1$).

\begin{figure}[htbp]
  \centering
  \includegraphics[width=7.5cm]{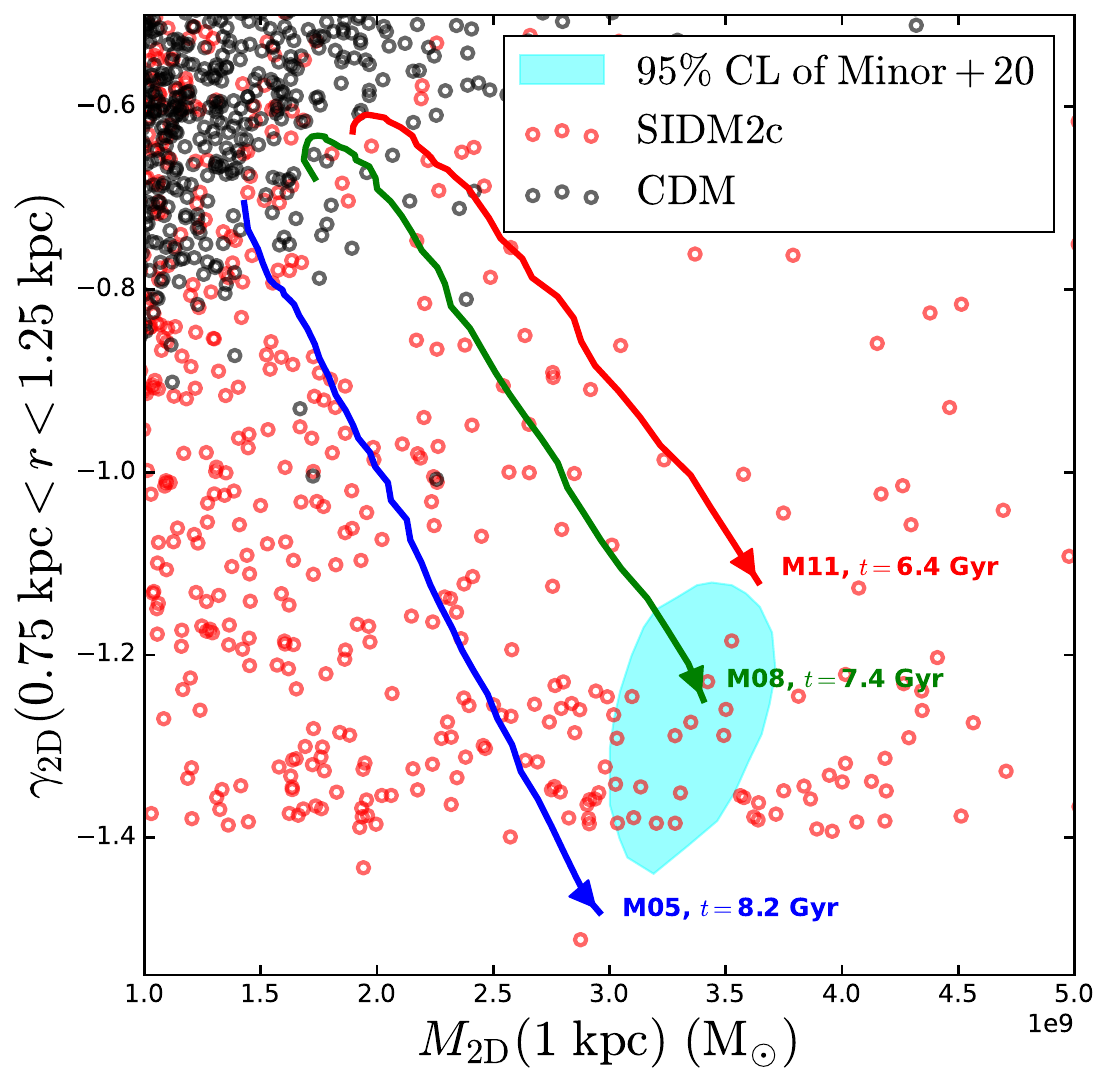}
  \caption{\label{fig:gamma} 
Projected logarithmic density slope $\gamma_{\rm 2D}$ (averaged over $0.75$--$1.25~\mathrm{kpc}$) vs. projected mass within 1 kpc for benchmark halos: \texttt{M11} ($10^{11}~\rm M_\odot$, red), \texttt{M08} ($8 \times 10^{10}~\rm M_\odot$, green), and \texttt{M05} ($5 \times 10^{10}~\rm M_\odot$, blue), all with $+2.5\sigma$ concentration. Cyan region denotes the $95\%$ confidence contour enclosed for the SDSSJ0946+1006 lensing perturber from Ref.~\cite{minor:2020hic}. Arrows trace the evolution of each halo under \texttt{SIDM2c}, which naturally drives systems into the favored region within a Hubble time. 
For comparison, we also show subhalos of the cluster host halo in the \texttt{SIDM2c} model. Their density profiles are obtained using the \texttt{SIDM2c} parametric model, whose functional form limits $\gamma_{\mathrm{2D}} \gtrsim -1.5$.
}
\end{figure}

\begin{figure*}[htbp]
  \centering

  \begin{overpic}[width=5.5cm]{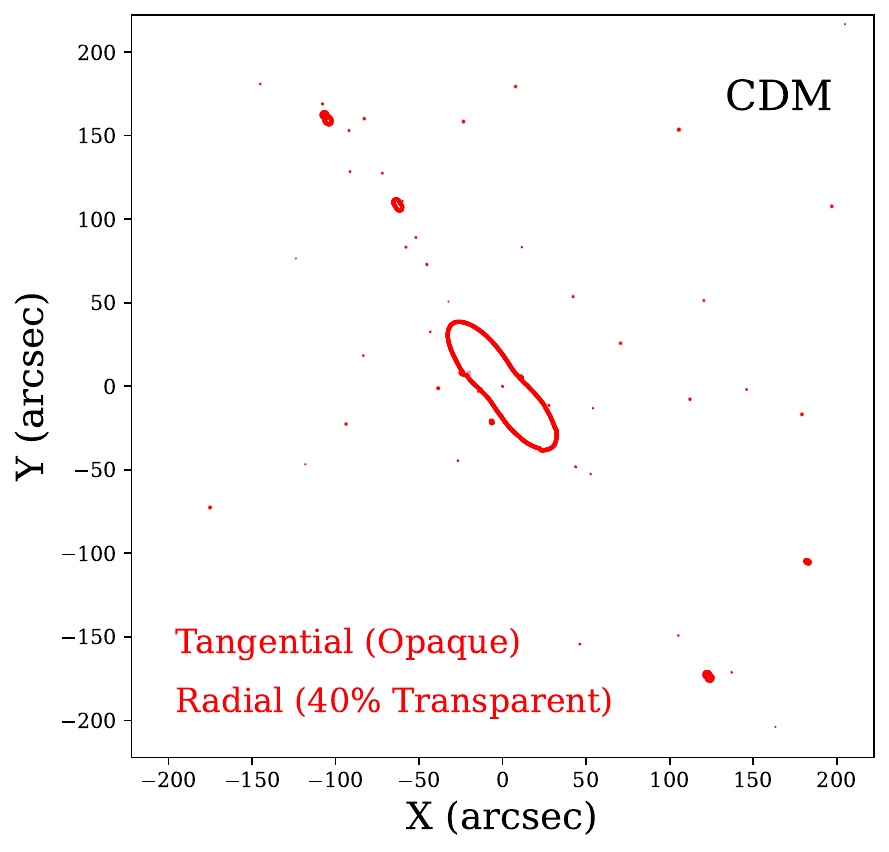}
    \put(2.,92){\textsf{\textbf{(a)}}}
  \end{overpic}
  \begin{overpic}[width=5.5cm]{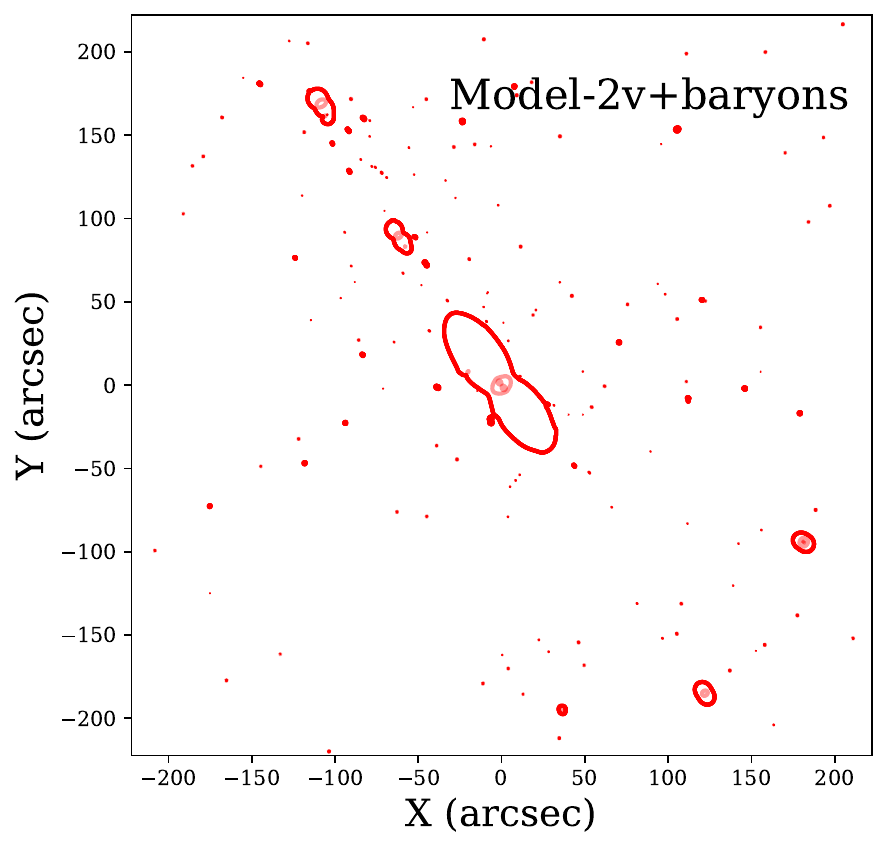}
    \put(2.,92){\textsf{\textbf{(b)}}}
  \end{overpic}
  \begin{overpic}[width=5.5cm]{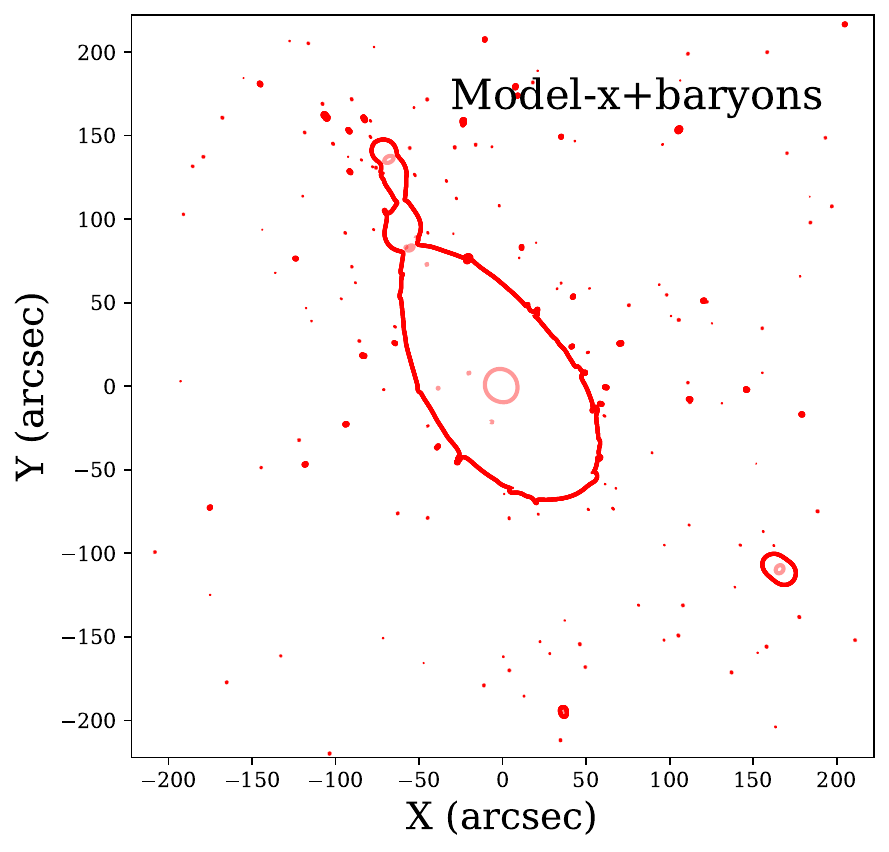}
    \put(2.,92){\textsf{\textbf{(c)}}}
  \end{overpic}

  \vspace{2mm}

  \begin{overpic}[width=5.5cm]{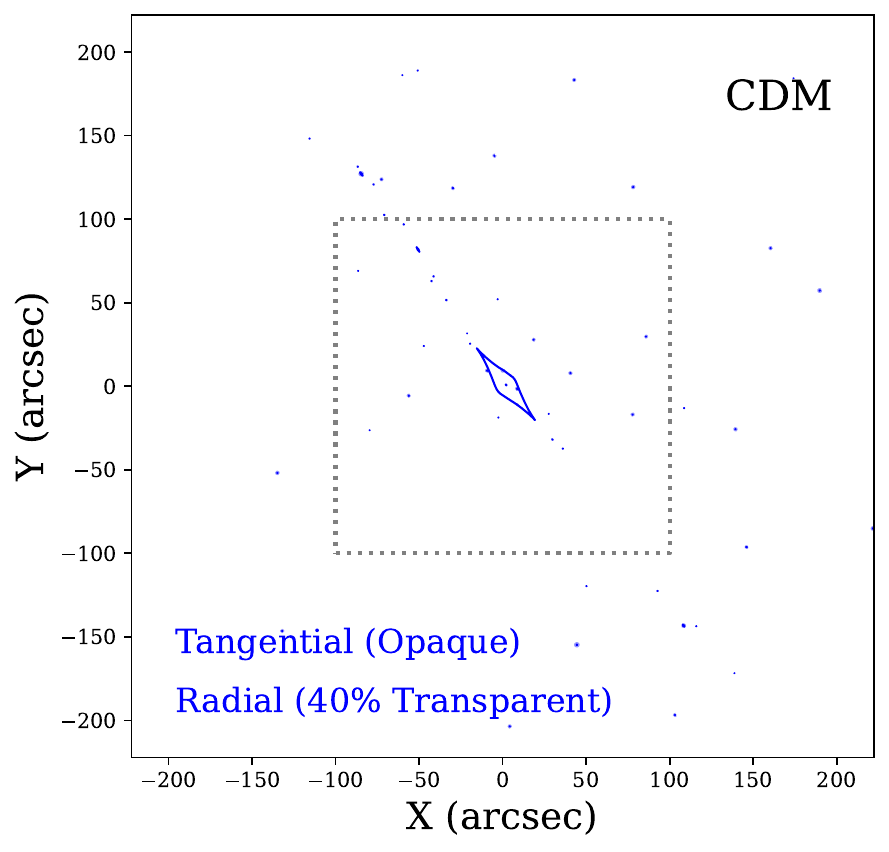}
    \put(2,92){\textsf{\textbf{(d)}}}
  \end{overpic}
  \begin{overpic}[width=5.5cm]{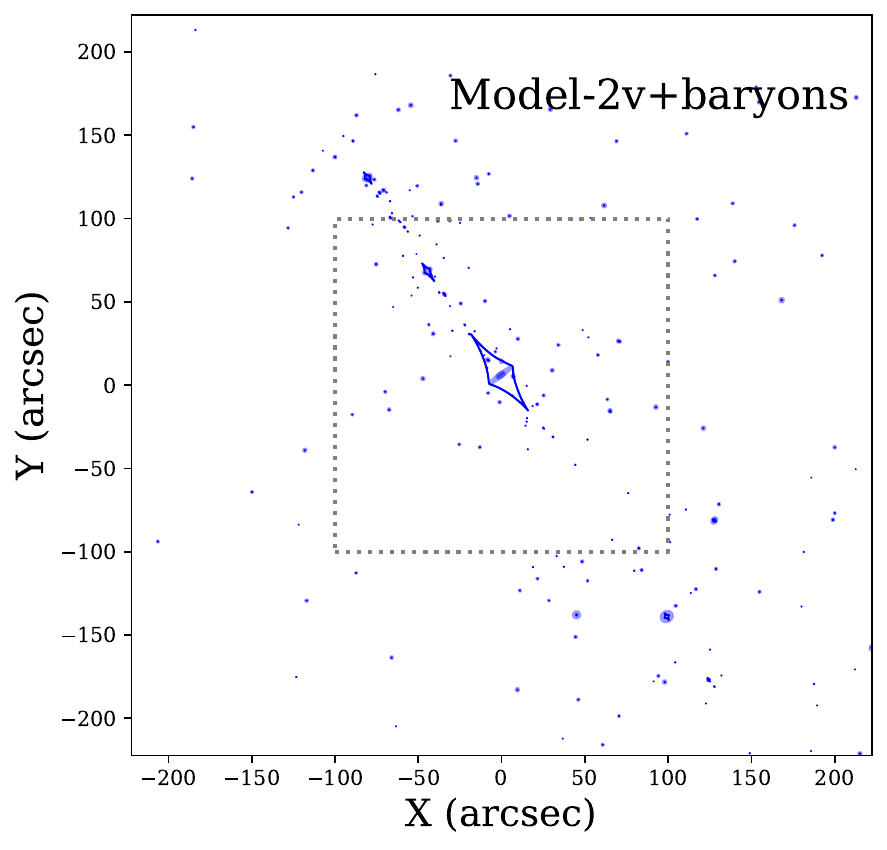}
    \put(2,92){\textsf{\textbf{(e)}}}
  \end{overpic}
  \begin{overpic}[width=5.5cm]{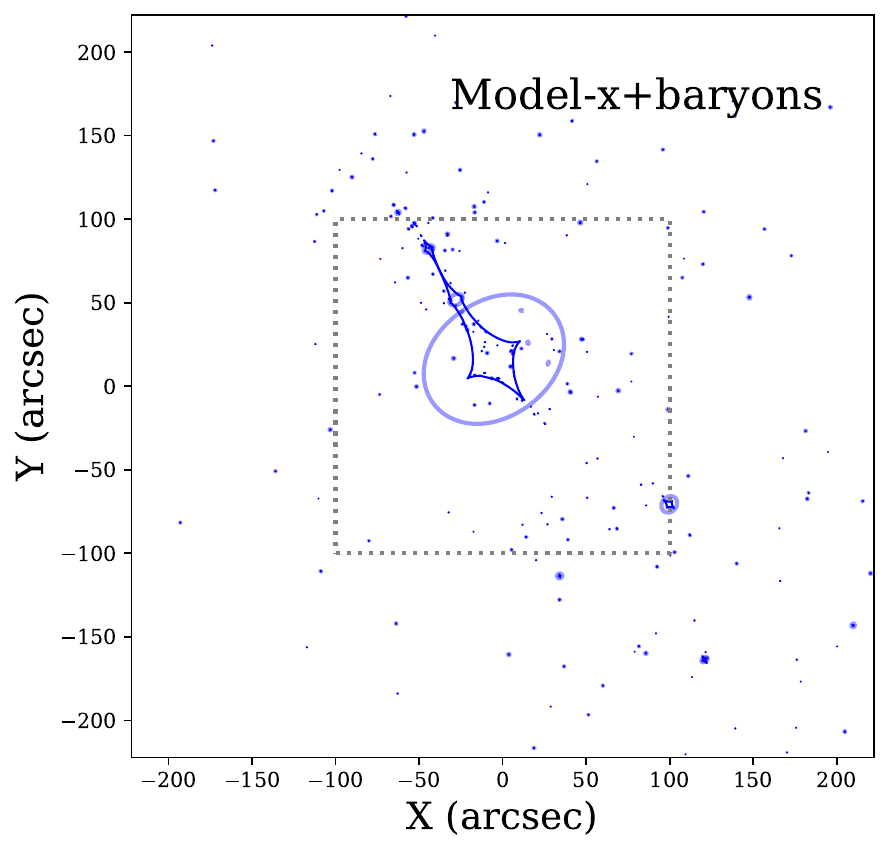}
    \put(2,92){\textsf{\textbf{(f)}}}
  \end{overpic}

  \caption{\label{fig:ggsl} 
Comparison of simulated gravitational lenses in \texttt{CDM} (a,d), \texttt{Model-2v} (b,e), and \texttt{Model-x} (c,f). Top panels show tangential (opaque) and radial (40\% transparent) critical curves overlaid on projected density maps; bottom panels show corresponding caustics. The two-component models produce significantly more secondary caustics due to dense substructures. 
The GGSL cross section ($\sigma_{\rm GGSL}$) is computed summing over the areas enclosed by the secondary caustics whose effective Einstein radii are larger than $0.5''$ and reside within the $(-100'',100'')$ square region centered on the host halo. 
}
\end{figure*}

\section{5. S\MakeLowercase{trong lensing perturbers}}

We have seen that mass segregation can produce higher inner halo densities than one-component SIDM, even during core formation. A denser inner halo also accelerates the halo’s gravothermal evolution. This is intuitive from the dependence of the core-collapse time $t_{\rm c}$ on halo concentration $c$.
A dimensional analysis of $t_{\rm c}$, translating the NFW parameters into mass and concentration, yields $t_{\rm c} \propto c^{(a-7)/2}$ for a generic $v^{-a}$ cross section in one-component SIDM. For both a velocity-independent case ($a=0$) and the commonly used Coulomb-like scattering ($a=4$), $t_{\rm c}$ decreases sharply as $c$ grows.
Hence, a more compact halo is expected to have a shorter $t_{\rm c}$.

Quantitatively, we calibrate the core collapse time of halos simulated in the \texttt{SIDM2c} model and find that it is close to that of a one-component SIDM model with $\sigma/m \approx 14~\text{cm}^2/\text{g}$, rather than the $0.3~\text{cm}^2/\text{g}$ value discussed earlier.
As shown in Fig.~\ref{fig:tau}, a small population of high concentration dwarf halos can evolve into dense, collapsing configurations under \texttt{SIDM2c}, explaining the observed dense dark substructures in strong lensing systems, commonly referred to as strong lensing perturbers~\cite{Vegetti:2008eg,2010MNRAS.408.1969V,2012Natur.481..341V,Hezaveh:2016ltk,minor:2020hic,Tajalli:2025qjx,Kong:2025sqx,He:2025wco,Lei:2025pky,Powell:2025rmj}.

Observationally, SDSSJ0946+1006 provides a pronounced example of a compact, dark substructure revealed through strong lensing, with image reconstruction indicating a steep projected density slope, $\gamma_{\rm 2D} \sim -1.25$, and high mass within the inner kiloparsec~\cite{minor:2020hic}. Such subhalos challenge both CDM and single-component SIDM models.
In CDM, it is particularly challenging to generate sufficiently dense and compact subhalos without invoking rare statistical fluctuations. While core-collapsing SIDM models can, in principle, produce such dense objects, prior studies have shown that cross sections of $\gtrsim 100~\rm cm^2/g$ are required for a $10^{11}\,M_{\odot}$ halo to reach collapse~\cite{nadler:2023nrd,2025ApJ...994..201L}, creating a potential tension with the much smaller cross sections inferred from dwarf galaxy clustering. This tension could be alleviated by incorporating a $v^{-4}$ velocity-dependence into the SIDM model, which would allow lower-mass halos to have larger cross sections, enabling some $10^{9}~\mathrm{M}_{\odot}$ halos to undergo core collapse.

To illustrate how SIDM with mass segregation can alleviate these tensions, we simulate three benchmark halos under \texttt{SIDM2c} with masses $M = 5 \times 10^{10}$, $8 \times 10^{10}$, and $10^{11}~\rm M_{\odot}$, labeled \texttt{M05}, \texttt{M08}, and \texttt{M11}, respectively. Each is initialized with a concentration $2.5\sigma$ above the cosmological median.

Fig.~\ref{fig:gamma} shows the evolution of the benchmark halos in the plane of projected inner mass (within 1 kpc) versus average projected slope $\gamma_{\rm 2D}$, with arrows indicating their trajectories over time. 
All three evolve toward the region favored by the SDSSJ0946+1006 perturber, with \texttt{M08} passing through the 95\% confidence contour at $t \approx 7.4$ Gyr. This demonstrates that mass segregation in two-component SIDM can drive certain halos into dense, lensing-relevant configurations even at modest cross sections.
By contrast, in \texttt{SIDM03}, the same halos remain in the core-forming phase and evolve away from the dense regime required to match the perturber.
To connect with the cosmological context, we overlay subhalos of the cluster host in \texttt{SIDM2c}, obtained using the parametric model. These subhalos extend to lower slopes and higher inner masses, with several candidates residing within the region favored by the SDSSJ0946+1006 perturber, supporting the viability of forming compact, dark structures in SIDM models with mass segregation.

\section{6. S\MakeLowercase{mall-scale lenses and lensing cross sections}}

The capability of enhancing inner masses in two-component models makes subhalos more efficient lenses and addresses the challenge reported in Ref.~\cite{Meneghetti:2020yif}, where the GGSL cross section, defined as the area enclosed by tangential caustics induced by subhalos, was initially found to be an order of magnitude larger in observations.
Refs.~\cite{Meneghetti:2022apr,Meneghetti:2023fug} further explored the effects of simulation resolution, AGN feedback, and galaxy formation models, etc., finding that the observed GGSL cross section remains a persistent factor of 3–6 times higher than predicted. 
This challenge remains difficult to reconcile with current models. Simulations with baryonic physics can increase subhalo densities, but the effect is often driven by one or two massive satellites and fails to robustly match observed statistics~\cite{2021MNRAS.505.1458B,2021MNRAS.504L...7R,Dutra:2024qac}. In SIDM, the core collapse has demonstrated potential, but achieving it across all subhalos with inner density slopes steeper than $r^{-2.5}$ remains challenging~\cite{Dutra:2024qac}. Moreover, when baryons are included, core-collapsed subhalos tend to enhance radial caustics rather than the tangential ones relevant for four-image systems~\cite{yang:2021kdf}.
In contrast, mass segregation naturally increases central densities, raising the enclosed mass above the critical surface density and enlarging the Einstein radii of SIDM halos. 

To quantify our framework's capability to reconcile the GGSL anomaly, we perform cosmological simulations for \texttt{CDM}, \texttt{SIDM2v}, and \texttt{SIDMx}. The \texttt{CDM} simulation assumes a one-component model, while the latter two are detailed in Table~\ref{tab:model}. 
The resolution of these simulations is insufficient to make accurate predictions for dwarf halos. Therefore, we adopt a hybrid approach. We use simulated halo density profiles only for candidates with masses above $10^{13}~\rm M_{\odot}$ in the \texttt{CDM} simulation. These massive halos are matched across all three simulations, and only matched halos are retained, ensuring that the large-scale densities remain consistent with the respective cosmological runs. 
For lower-mass halos, we apply the parametric \texttt{SIDM2c} model to the corresponding \texttt{CDM} halos. 
We adopt the integral approach from Ref.~\cite{yang:2023jwn} to obtain predictions that incorporate the effects of accretion histories. This approach provides results that differ from the simulated \texttt{SIDM2v} and \texttt{SIDMx} models.
For clarity, we refer to these hybrid models as \texttt{Model-2v} and \texttt{Model-x}, corresponding to cases where the massive halos are taken from the \texttt{SIDM2v} and \texttt{SIDMx} simulations, respectively. 
In effect, our study focuses on the impact of \texttt{SIDM2c} halos in the mass range $10^{10}$--$10^{12}~\rm M_{\odot}$ on GGSL. The integrated results are presented for two alternative models: one representing a scenario that remains safe from cluster SIDM constraints (\texttt{Model-2v}), and the other exhibiting extreme mass segregation (\texttt{Model-x}).
We also incorporate the effect of baryons on halo density profiles using a parametric approach, obtaining results for the \texttt{Model-2v+baryons} and \texttt{Model-x+baryons} scenarios. 
For consistency with the CDM comparisons, the baryon potentials themselves are not included directly in the lensing analysis. 

To facilitate comparison with the GGSL measurement in Ref.~\cite{Meneghetti:2020yif}, we fix the redshift of our host system at $z = 0.439$ and compute lensing properties by including all subhalos within a $5~{\rm Mpc}/h$ sphere.
The lensing potential and deflection angles are primarily modeled analytically using the parametric SIDM halo profile developed in Refs.~\cite{Hou:2025gmv,yang:2023jwn}, which allows for efficient calculation of lensing observables across a wide range of halo masses and internal structures. See the Supplemental Material for details~\cite{balberg:2002ue,1994A&A...284..285K,2013mnras.428.3121m,2019MNRAS.485..382C}. 

Fig.~\ref{fig:ggsl} compares the tangential (opaque) and radial (40\% transparent) critical curves (Fig.~\ref{fig:ggsl}a, b, and c) and associated caustics (Fig.~\ref{fig:ggsl}d, e, and f) in the \texttt{CDM}, \texttt{Model-2v}, and \texttt{Model-x} simulations. 
We observe a substantial increase in the number and size of small-scale caustics in both two-component SIDM models compared to CDM. 

Quantitatively, we compute the fiducial GGSL cross section ($\sigma_{\rm GGSL}$, {in unit of arcsec$^{2}$}) as the total area enclosed by secondary caustics (excluding the host caustic), within a $(-100'', 100'')$ square region centered on the host halo. Following Meneghetti et al. \cite{Meneghetti:2020yif}, we apply a lower area threshold of $\pi(0.5'')^2$. 
We also exclude any contribution with effective Einstein radius $\theta_E\equiv\sqrt{A_c/\pi}$ being larger than three arcseconds, where $A_c$ is the area enclosed by a critical line.
This cut prevents the statistical dominance of massive outliers, as they are physically distinct from dwarf halos and are not observed to produce $\theta_E>3''$~\cite{Meneghetti:2022apr,Meneghetti:2023fug}. 
We repeated the lensing analysis 11 times, corresponding to 11 different viewing angles. Each viewing angle was generated by consecutively rotating the system around the $x$- and $y$-axes in steps of $18^\circ$, i.e., $18^\circ$, $36^\circ$, $54^\circ$, $72^\circ$, and $90^\circ$ rotations for each axis, plus a case without rotation.

\begin{table}[h]
\centering
\caption{GGSL cross section comparisons. }
\label{tabA1}
\begin{tabular}{lcccc}
\hline
Model & Mean GGSL & Std Dev & Total N & Ratio \\
\hline
CDM & 0.3024& 0.2484& 41 & 1.00 \\
Model-2v & 0.6133& 0.3689& 85 & 2.03 \\
Model-x & 2.223& 0.7615& 207 & 7.35 \\
\hline
Model-2v+baryons & 1.360& 0.5023& 191 & 4.50 \\
Model-x+baryons & 4.149& 1.602& 326 & 13.7 \\
\hline
\end{tabular}
\vspace{0.5em}
\parbox{\linewidth}{
\footnotesize
The mean GGSL cross sections with their standard deviations and total number of contributing secondary caustics (aggregated over 11 analyses) are reported for different model scenarios. The ratio column computes the ratio of mean GGSL relative to the value in CDM, which illustrates the amount of excess. The GGSL cross section is computed considering secondary critical lines with $\theta_E<3''$.
}
\end{table}

Table~\ref{tabA1} summarizes the results of our GGSL analysis. The table presents, for each simulation, the mean GGSL cross section, its standard deviation, and the total number of contributing secondary caustics (aggregated over 11 analyses). 
The final column reports the ratio of the mean GGSL cross section relative to the CDM value, quantifying the amount of excess produced by two-component SIDM models. All values are computed considering only secondary critical lines with $\theta_E < 3''$. 
The \texttt{Model-2v} and \texttt{Model-x} benchmarks correspond to host halos that are shallower and denser, respectively, than the CDM host halo. As expected, the GGSL cross section is higher in the denser \texttt{Model-x} case. In \texttt{Model-2v}, the GGSL cross section and the number of contributing caustics are both approximately double the CDM values. This enhancement increases to about 4.5 times the CDM value when baryons are included. The excess is systematically larger in \texttt{Model-x}, where the ratio rises to 7.35 without baryons and 13.7 with baryons.
The level of GGSL excess produced by our benchmark models demonstrates their potential to address the observed GGSL excess, which is a factor of 3-6 above theoretical expectations.
Beyond the total GGSL cross section, we also compare the distribution of Einstein radii of the contributing secondary critical lines with observations. Additional details are provided in Appendices D and E.

\section{7. C\MakeLowercase{onclusion}}

Observations on small scales reveal the simultaneous presence of cored dwarf galaxies and dense substructures in host halos spanning multiple mass scales. This study demonstrates that such features naturally coexist in SIDM models with mass segregation. 
We also reveal a conditioned universality in two-component SIDM models, which we use to construct a parametric model for generating theoretical predictions for a population of halos.

Future work could improve upon several aspects of this work, including the simulation resolution, the scan of model parameters, and the incorporation of baryonic physics. 
While controlled simulations are used to illustrate the effect of baryons, more dedicated simulations are required to map the full landscape of baryonic effects in \texttt{SIDM2c} halos. 
In the present analysis, resolution limitations are mitigated using a parametric modeling approach, and the analysis is based on a single cluster zoom evaluated across 11 projections, which limits direct comparison with observations. Additional uncertainties can arise from host-dependent properties.

Despite these limitations, the results support SIDM with mass segregation as a plausible framework for addressing multiple small-scale challenges.
Looking ahead, our framework may also shed light on the formation of early massive galaxies and supermassive black holes, including the so-called ``little red dots'', as mass segregation remains effective at high redshifts. Our results suggest that small-scale structure may already be revealing not only the strength but also the multi-component nature of dark matter. \\

\textbf{Conflict of interest} \\

The authors declare that they have no conflict of interest.\\

\textbf{Acknowledgements} \\

We thank Moritz Fischer, Guoliang Li, Nan Li, Ran Li, Shihong Liao, Massimo Meneghetti, Priyamvada Natarajan, Huiyuan Wang, Hai-Bo Yu, and Yi-Ming Zhong for helpful discussion. 
The authors were supported in part by the National Key Research and Development Program of China (2022YFF0503304), Natural Science Foundation of China (12588101), Chinese Academy of Sciences (CAS), and the Project for Young Scientists in Basic Research of CAS (YSBR-092). \\

\textbf{Author contributions} \\

Daneng Yang and Yi-Zhong Fan proposed this study. Daneng Yang led the analysis and performed the simulations. Daneng Yang, Yi-Zhong Fan, and Yue-Lin Sming Tsai collaborated on writing the manuscript. Siyuan Hou contributed in part to the GGSL analysis. Yue-Lin Sming Tsai and Yi-Zhong Fan acquired funding for the analysis, particularly the simulations. \\

%

\appendix
\setcounter{figure}{0}
\renewcommand{\thefigure}{S\arabic{figure}}
\renewcommand{\figurename}{Fig.}

\setcounter{table}{0}
\renewcommand{\thetable}{S\arabic{table}}
\renewcommand{\tablename}{Table}

\setcounter{equation}{0}
\renewcommand{\theequation}{S\arabic{equation}}


\begin{figure*}[htbp]
  \centering
  \begin{overpic}[width=5.5cm]{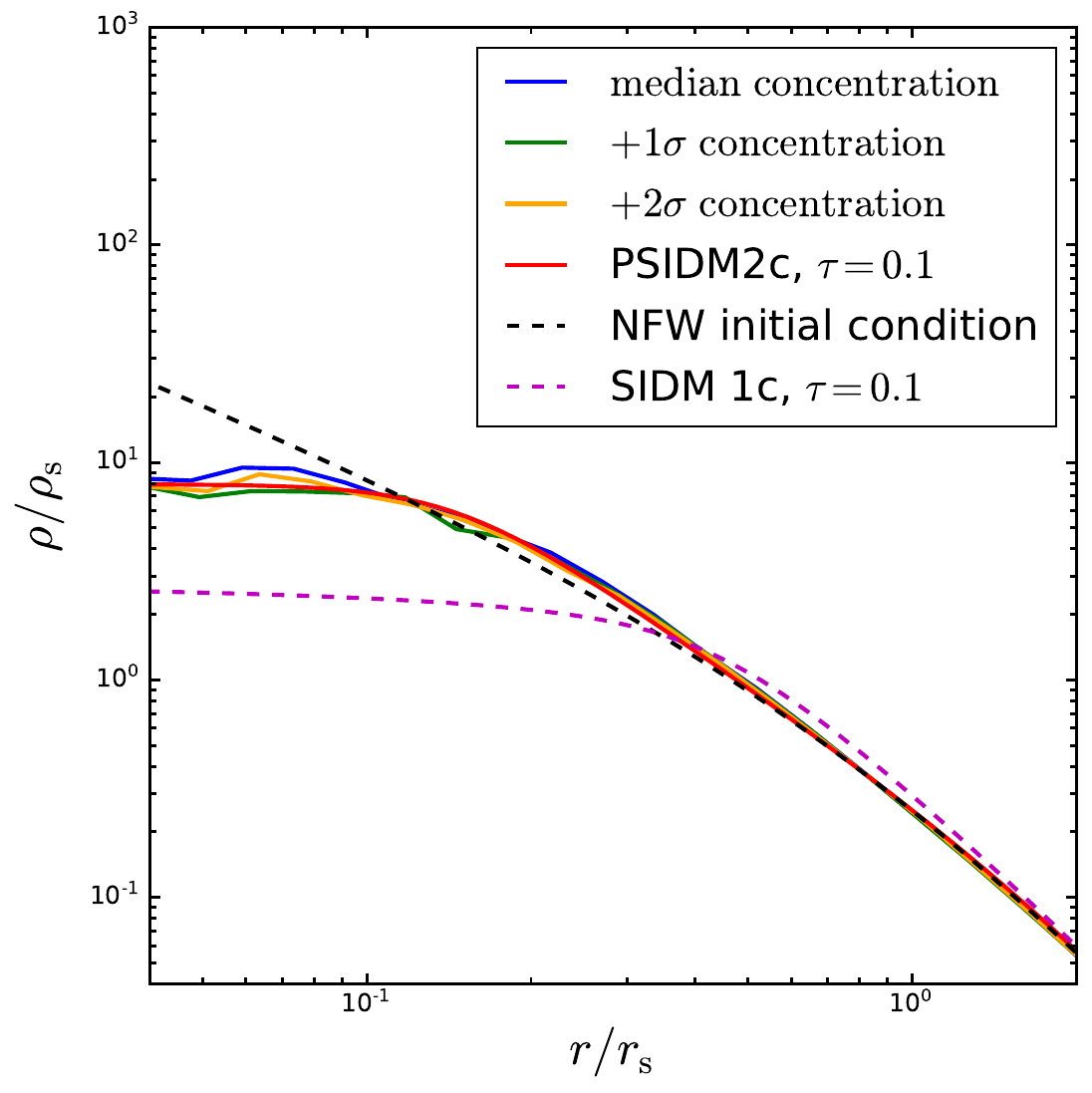}
    \put(2,92){\textsf{\textbf{(a)}}}
  \end{overpic}
  \begin{overpic}[width=5.5cm]{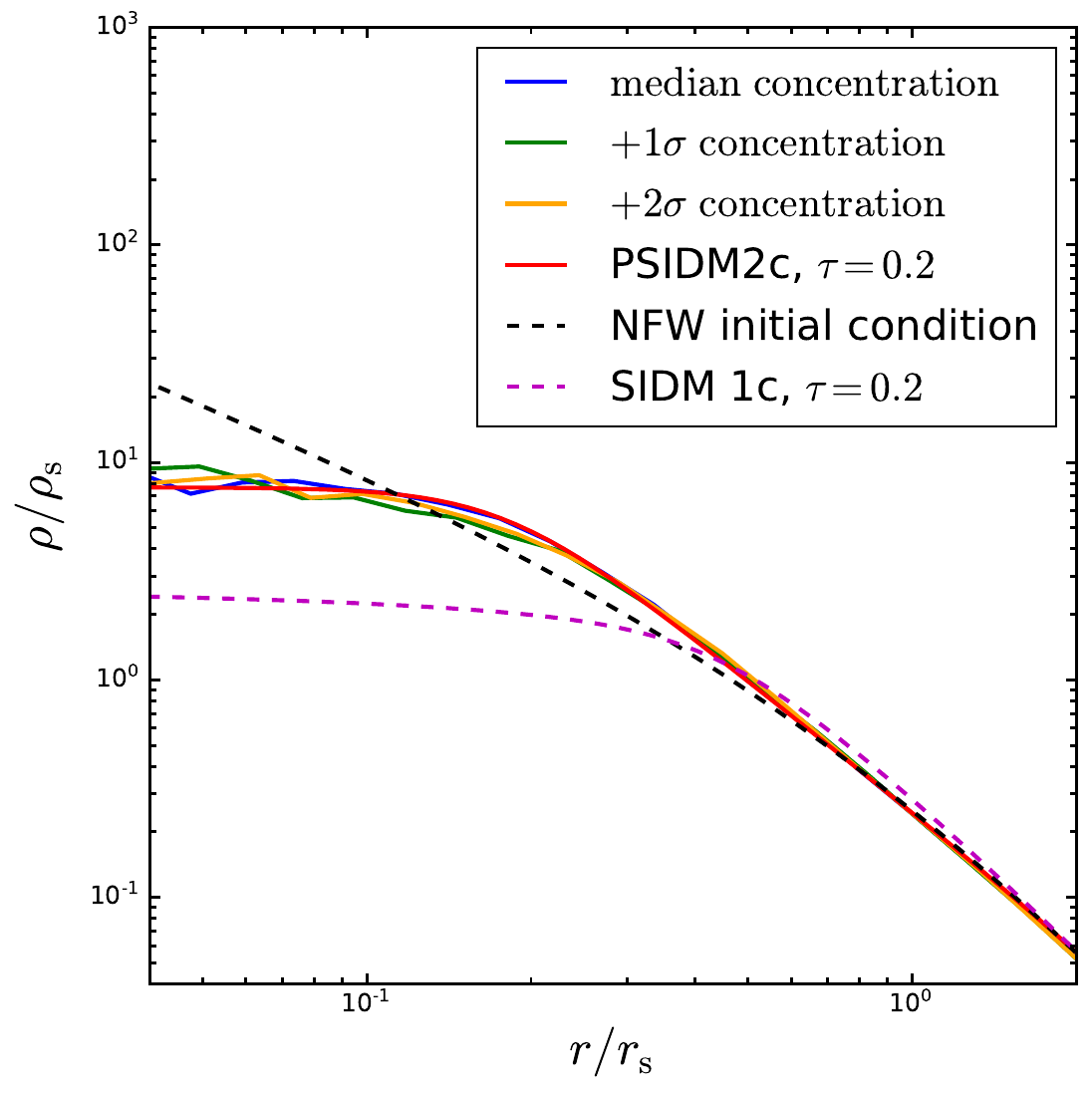}
    \put(2,92){\textsf{\textbf{(b)}}}
  \end{overpic}
  \begin{overpic}[width=5.5cm]{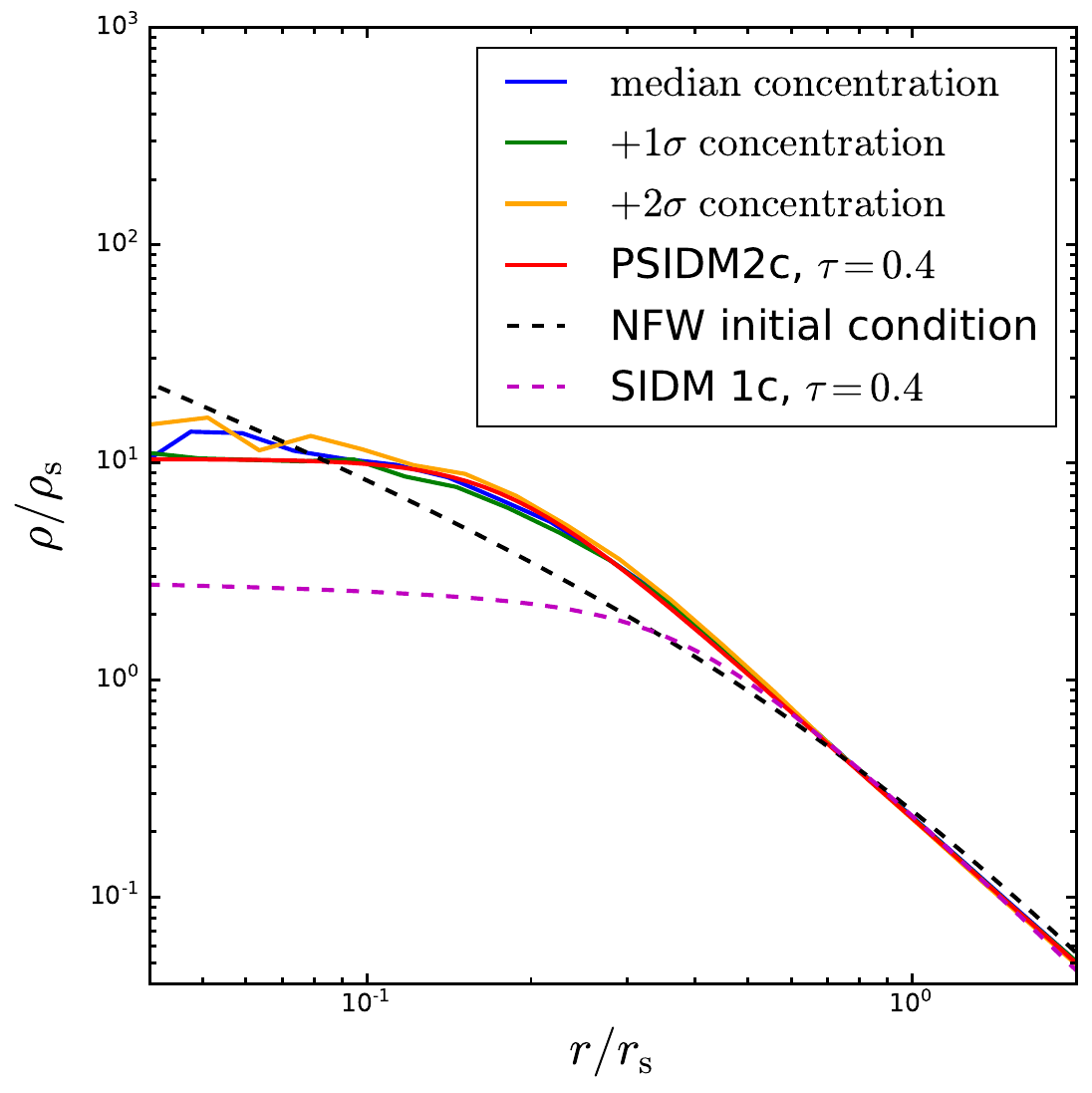}
    \put(2,92){\textsf{\textbf{(c)}}}
  \end{overpic}

  \vspace{2mm}

  \begin{overpic}[width=5.5cm]{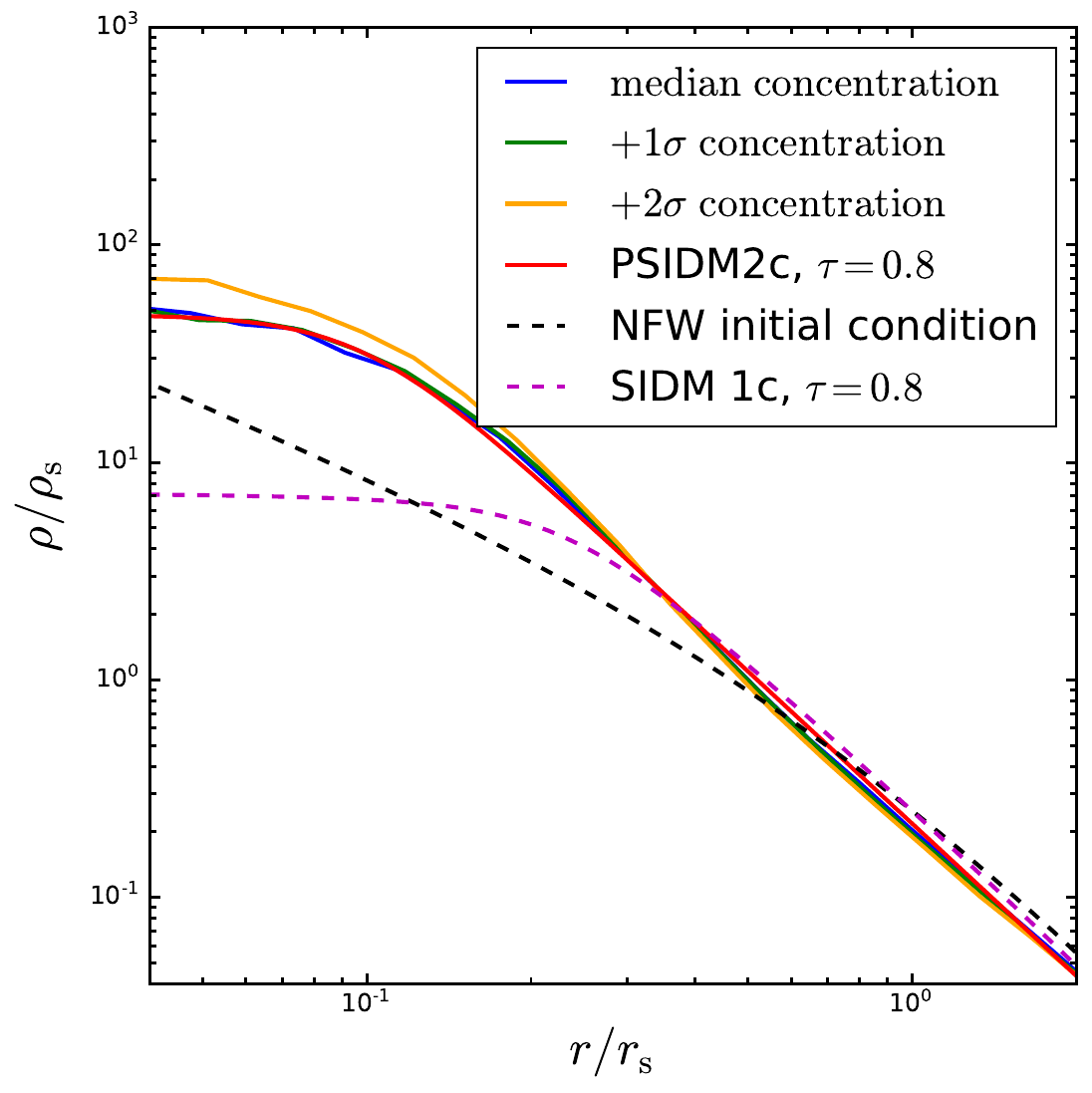}
    \put(2,92){\textsf{\textbf{(d)}}}
  \end{overpic}
  \begin{overpic}[width=5.5cm]{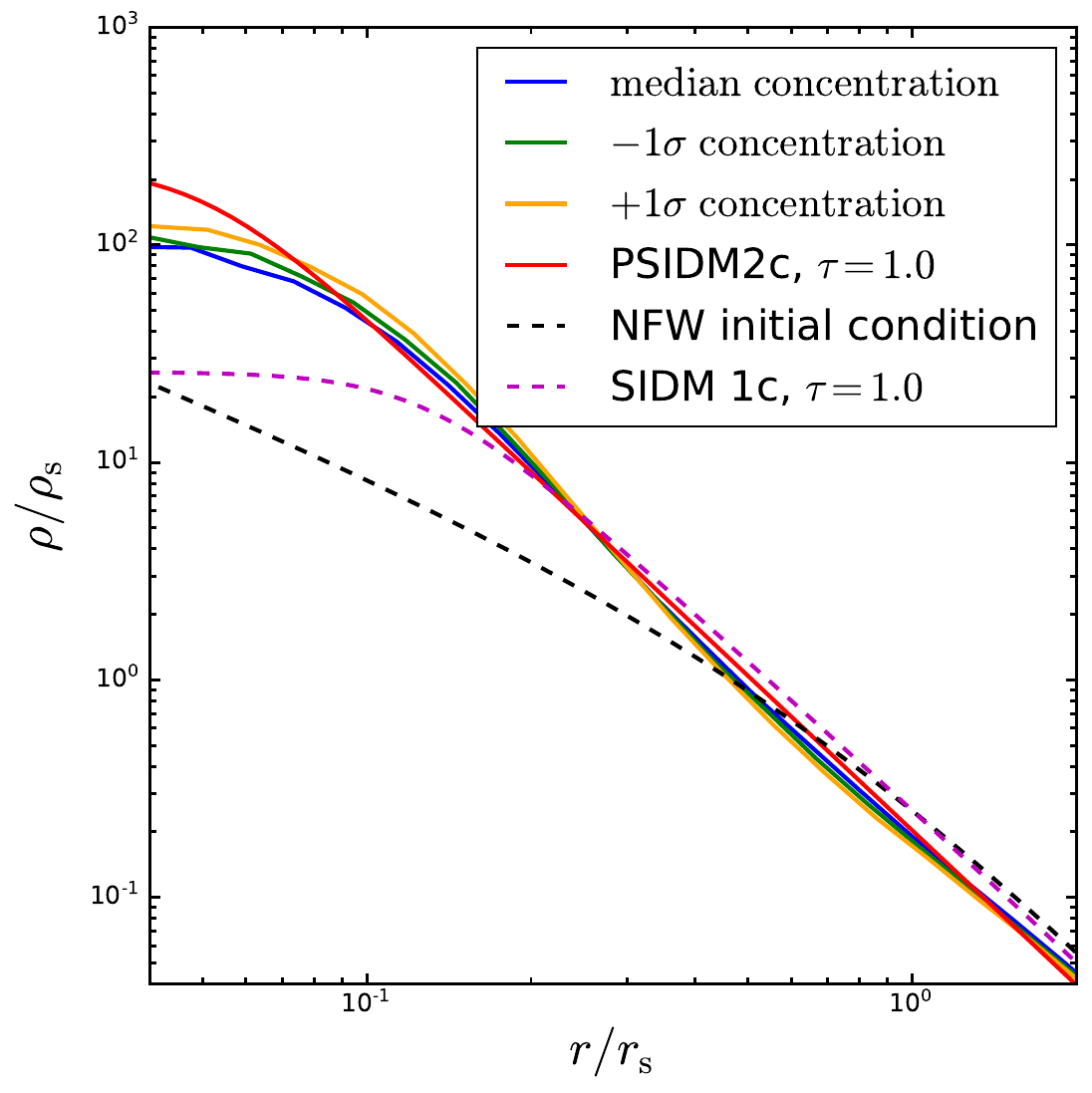}
    \put(2,92){\textsf{\textbf{(e)}}}
  \end{overpic}
  \begin{overpic}[width=5.5cm]{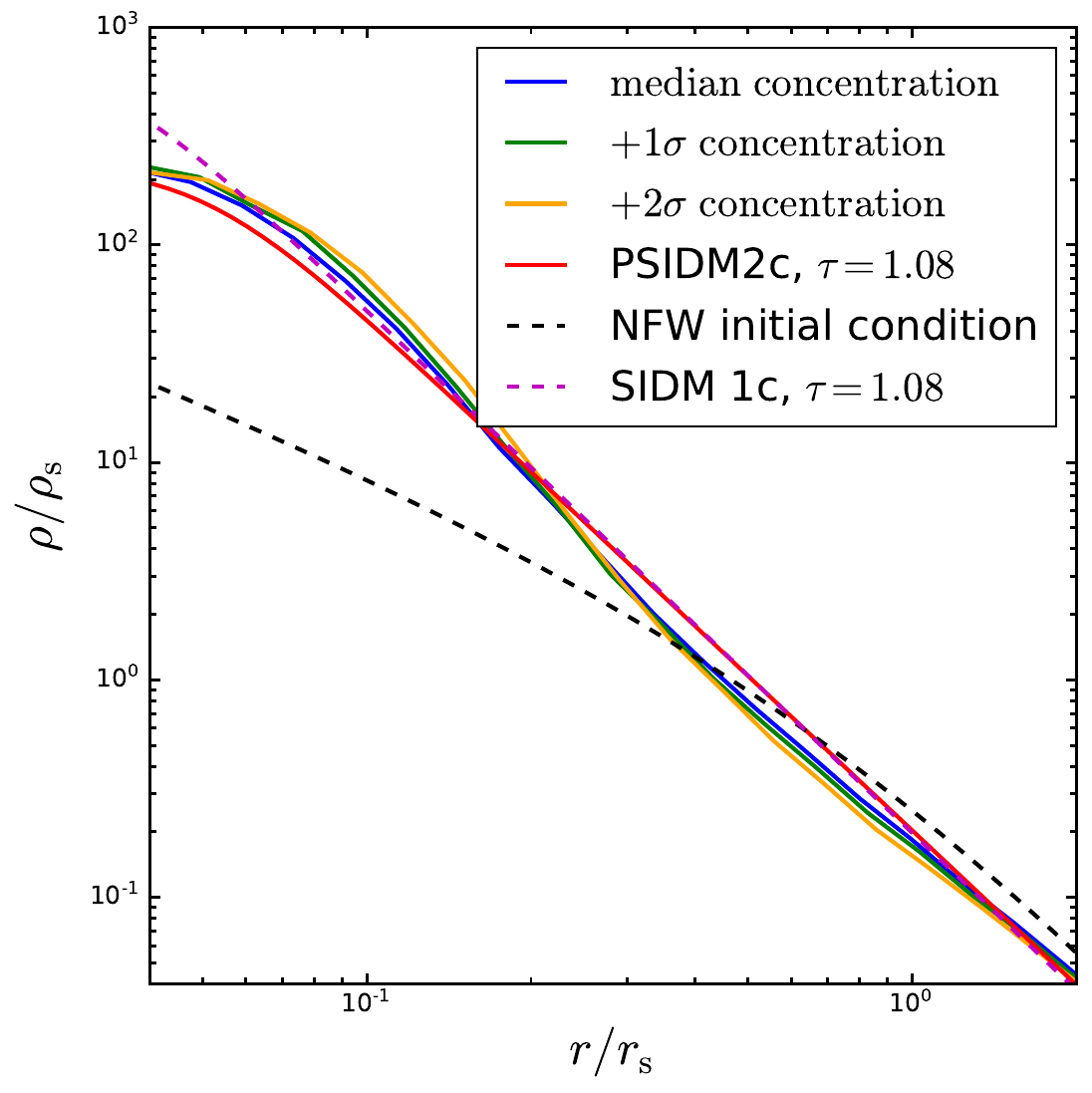}
    \put(2,92){\textsf{\textbf{(f)}}}
  \end{overpic}

  \caption{\label{fig:normprofs} Normalized density profiles for three $2\times 10^{10}~\rm M_{\odot}$ halos with median (blue), $+1\sigma$ (green), and $+2\sigma$ (orange) concentrations in the \texttt{SIDM2c} model, shown at increasing values of the gravothermal phase $\tau$. For comparison, the magenta curves show corresponding one-component SIDM profiles, and the initial NFW profile is shown in black. The red curves show the parametric model predictions for \texttt{SIDM2c}. The close match with the simulated curves demonstrates the quality of our calibration. The agreement among the three \texttt{SIDM2c} profiles (solid curves) in each panel demonstrates the conditioned universality of the model. Notably, the inner densities in the two-component case are consistently higher than in the one-component case apart from the late gravothermal phases ($\tau > 1$), indicating that SIDM with mass segregation is more effective at enhancing the strong lensing signal.
}
\end{figure*}

\begin{figure*}[htbp]
  \centering

  \begin{overpic}[width=5.5cm]{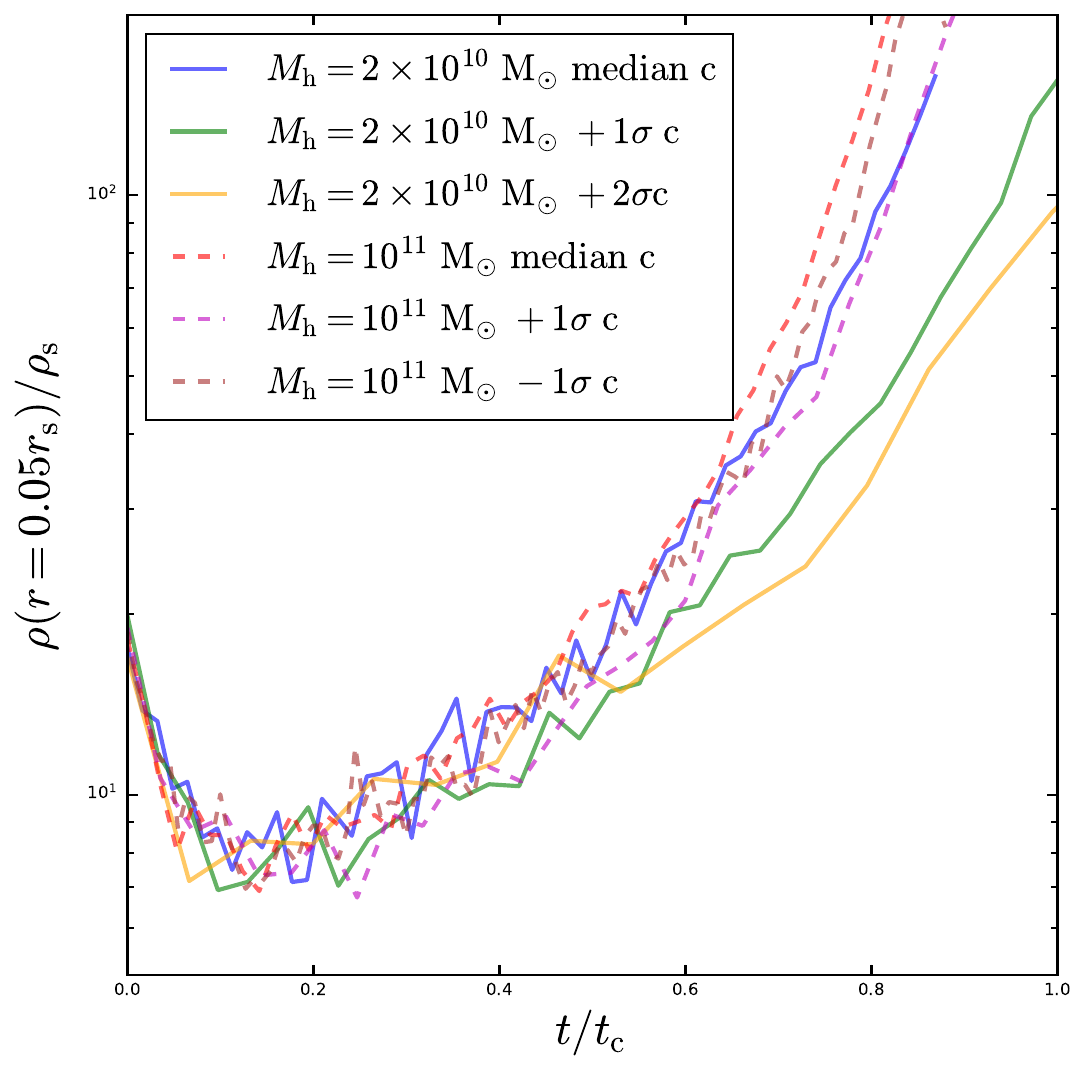}
    \put(2,92){\textsf{\textbf{(a)}}}
  \end{overpic}
  \begin{overpic}[height=5.5cm]{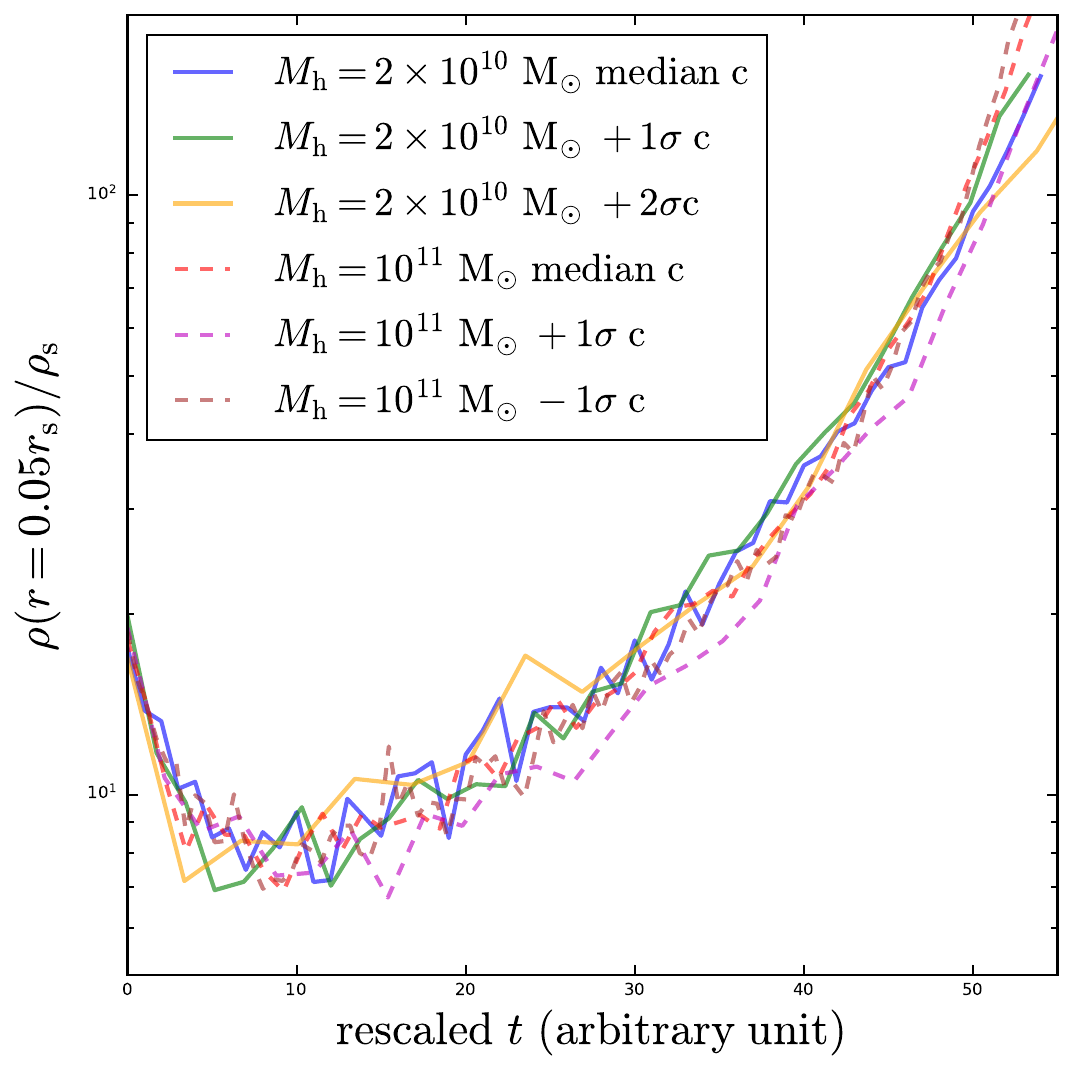}
    \put(2,92){\textsf{\textbf{(b)}}}
  \end{overpic}
  \begin{overpic}[height=5.5cm]{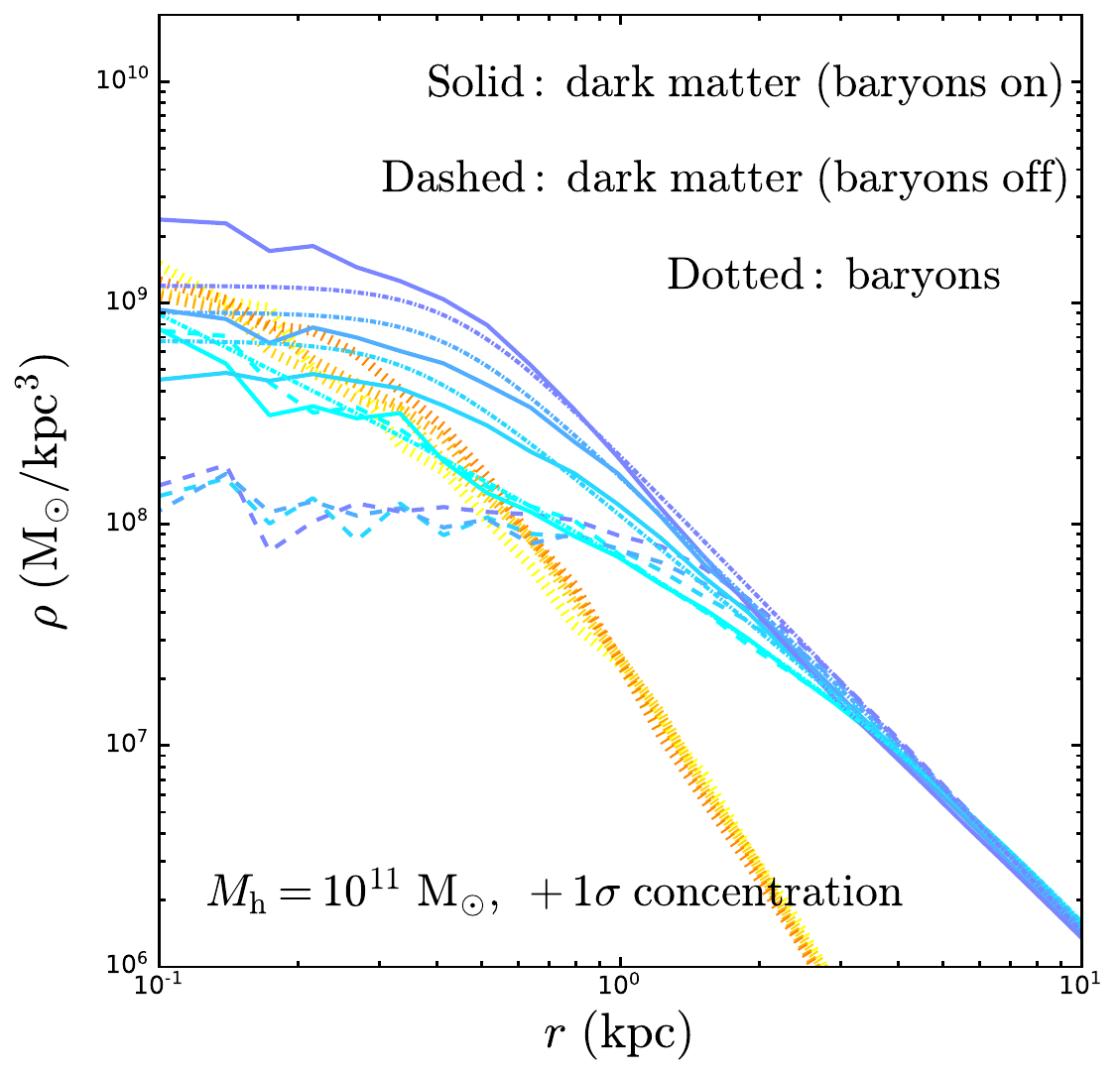}
    \put(1,88){\textsf{\textbf{(c)}}}
  \end{overpic}

  \caption{\label{fig:rhoctc} \textbf{(a): } Normalized inner halo density evolution in \texttt{SIDM2c} for halos of $2\times10^{10}~M_{\odot}$ and $10^{11}~M_{\odot}$ with various concentrations. The evolution time is normalized by the core collapse time calculated using Eq.~(\ref{eq:tc0}) with $\sigma/m=14~\rm cm^2/g$. 
  \textbf{(b): } As in the left, but with the evolution time of all the simulations normalized to the first case in blue. 
  \textbf{(c): } Model predictions (dash-dotted lines) versus simulation (shaded bands) for a $10^{11}~M_{\odot}$ halo ($+1\sigma$ concentration, with baryons) at $t = 0,3,6,9$ Gyr. Solid and dotted lines correspond to the halo and baryon profiles in the baryonic simulation, respectively. Dashed lines represent the halo profile in the DM-only simulation.  
}
\end{figure*}

\section{A. A parametric model for the two-component model SIDM2c}

\subsection{Verification of the conditioned universality}

In the main text, we illustrated the existence of a conditioned universality in SIDM models with mass segregation by showing the rescaled halo density profile at a normalized gravothermal phase of $\tau \equiv t/t_{\rm c} = 0.6$.  
Here, we further explore these universal features in Fig.~\ref{fig:normprofs} by showing density profiles at additional gravothermal phases: $\tau = 0.1, 0.2, 0.4, 0.8, 1.0,$ and $1.08$.  
For simplicity, we only show profiles for the three halos of mass $2 \times 10^{10}~\rm M_{\odot}$ and compare them with the one-component SIDM results (dashed magenta curves) at the same gravothermal phases. 
The one-component results are obtained using a parametric model calibrated to the conduction fluid model from Ref.~\cite{Hou:2025gmv}.

From Fig.~\ref{fig:normprofs}, it is evident that the inner halo densities in the \texttt{SIDM2c} model are consistently higher than those in the one-component SIDM case across almost all gravothermal phases.  
Interestingly, in the late gravothermal phases ($\tau > 1$), this difference diminishes, and the \texttt{SIDM2c} profiles become comparable to those of the one-component model.  
This result implies that the enhancement induced by \texttt{SIDM2c} does not stem from deeply core-collapsed halos, but rather from a broad population of halos across a range of gravothermal phases.

\subsection{A parametric model for DM-only halos in SIDM2c}

To make use of the conditioned universality, we construct a parametric model for the \texttt{SIDM2c} scenario based on the simulated $2 \times 10^{10}~\rm M_{\odot}$ halo with $+1\sigma$ concentration. We find the core collapse time in \texttt{SIDM2c} to be approximately that of a constant one-component SIDM model with $\sigma/m=14~\rm cm^2/g$. This allows us to compute the core collapse time as~\cite{balberg:2002ue,essig:2018pzq}:
\begin{eqnarray}
\label{eq:tc0}
t_{\rm c}  &=& \frac{150}{C} \frac{1}{(\sigma/m) \rho_{\rm s} r_{\rm s}} \frac{1}{\sqrt{4\pi G \rho_{\rm s}}},
\end{eqnarray}
where we adopt $C=0.75$, and $\rho_{\rm s}$ and $r_{\rm s}$ are the NFW scale parameters for the CDM halo under consideration.

Fig.~\ref{fig:rhoctc}a illustrates the evolution of the inner halo density, where time is normalized by the core collapse time $t_{\rm c}$ defined in Eq.~(\ref{eq:tc0}). 
For each simulation, the inner density is computed at $r=0.05 r_{\rm s}$ using the NFW scale parameters $\rho_{\rm s}$ and $r_{\rm s}$, and normalized by $\rho_{\rm s}$.
Without the time rescale by $t_{\rm c}$, the core collapse times for these halos range from 15 to 110 Gyr. 
A clear, unified evolutionary trend emerges after normalization, despite a roughly 30\% uncertainty in $t_{\rm c}$, which is comparable to the uncertainty in the one-component case. 

Note that the conditioned universality in theory is independent of the $t_{\rm c}$ formula. In Fig.~\ref{fig:rhoctc}b, we rescale the evolution time in all the cases to the core collapse time of the first case shown.
The agreement among the curves significantly improves over the left panel, hence future work can involve calibrating a more accurate expression for $t_{\rm c}$ to reduce uncertainty in the parametric model.

The parametric model is based on the density profile introduced in Ref.~\cite{Hou:2025gmv}, calibrated by fitting the evolution of its profile parameters to the simulated \texttt{SIDM2c} halo.

The model is parameterized as
\begin{eqnarray}
    \label{eq:PSIDM2c}
    &&\rho_{\rm PSIDM2c}(r|\rho_{s,0},r_{s,0},\tau) \\ \nonumber
    &&= \frac{\rho_{\rm s}}{\left[\left(\frac{r}{r_{\rm s}}\right)^4 + \left(\frac{r_c}{r_{\rm s}}\right)^4\right]^{\frac{\gamma}{4}} \cdot \left[1 + \left(\frac{r}{r_{\rm s}}\right)^\beta\right]^{\frac{3-\gamma}{\beta}}},\\ \nonumber
\end{eqnarray}
where $\rho_{\rm s}=\rho_{\rm s}(\rho_{s,0},\tau)$, $r_{\rm s}=r_{\rm s}(r_{s,0},\tau)$, $r_c=r_c(r_{s,0},\tau)$, $\beta=\beta(\tau)$ and $\gamma=\gamma(\tau)$ are functions of the phase parameter $\tau \in (0, 1.08)$. The profile parameters as functions of $\tau$ and are fitted to be

\begin{widetext}
\begin{eqnarray}
\label{eq:tracks}
\rho_{\rm s}(\tau)/\rho_{\rm s,0} &=& \exp\left(-15.74 \ln(3.748\tau + 1) - 0.4013 + 31.26\tau - 15.58\tau^2 + 2.999\tau^3\right), \\
r_{\rm s}(\tau)/r_{\rm s,0} &=& 1.116 + 7.066\tau + 38.89\tau^2 - 217.3\tau^3 + 484.3\tau^4 - 557.2\tau^5 + 328.1\tau^6 - 77.84\tau^7, \\
r_c(\tau)/r_{\rm s,0} &=& 0.5878\sqrt{\tau} - 0.6073\tau + 1.102\tau^{3/2} - 1.384\tau^2 - 1.492\tau^3 + 5.162\tau^4 - 4.951\tau^5 + 1.634\tau^6, \\
\beta(\tau) &=& 1.128 + 9.554\tau - 30.26\tau^2 + 124.9\tau^3 - 109.4\tau^4 - 181.9\tau^5 + 276.5\tau^6, \\
\gamma(\tau) &=& 0.2651 + 3.941\sqrt{\tau} + 0.2936\tau - 9.062\tau^2 + 14.35\tau^3 - 10.22\tau^4 + 2.785\tau^5 \\
&&\quad + \frac{1}{\ln(0.001)} \left[1 - 0.2651\right] \ln(\tau + 0.001).
\end{eqnarray}
\end{widetext}

Based on the parameterized density profiles, we also extract the evolution of $V_{\rm max}$ and $R_{\rm max}$, where $V_{\rm max}$ denotes the maximum rotation velocity and $R_{\rm max}$ its corresponding radius. We obtain the following fitting functions to describe their evolution:
\begin{widetext}
\begin{eqnarray}
\label{eqVR}
V_{\rm max,SIDM2c}(\tau)/V_{\rm max,0} &=& 1 - 0.2850\,\tau + 0.5936\,\tau^2 + 0.01723\,\ln(221.2\,\tau + 1) - 0.3175\,\tau^3 + 0.2157\,\tau^4, \\
R_{\rm max,SIDM2c}(\tau)/R_{\rm max,0} &=& 1 + [1 - (1 + e^{-20(\tau - 0.3)})^{-1}]P_1(\tau) + (1 + e^{-20(\tau - 0.3)})^{-1}P_2(\tau) \\
P_1(\tau) &=& -2.305\,\tau + 17.23\,\tau^2 - 45.64\,\tau^3 \\
P_2(\tau) &=& -0.7538\,\tau - 2.228\,\tau^2 + 3.508\,\tau^3 - 1.429\,\tau^4.
\end{eqnarray}
\end{widetext}

To map CDM halos from cosmological simulations to their \texttt{SIDM2c} counterparts, we incorporate the effect of accretion histories using the integral approach proposed in Ref.~\cite{yang:2023jwn}. In this approach, the gravothermal evolution is integrated along the evolutionary trajectories of $V_{\rm max,CDM}$ and $R_{\rm max,CDM}$ in the CDM simulation, yielding their predicted values in \texttt{SIDM2c} as
\begin{widetext}
\begin{eqnarray}
\label{eq:int}
\nonumber 
V_{\rm max}(t)    &=&  V_{\rm max, CDM}(t_f) + \int_{t_f}^{t} d t'  \frac{d V_{\rm max,CDM}(t')}{d t'} +  \int_{t_f}^{t} \frac{dt'}{t_{\rm c}(t')} \frac{d V_{\rm max, SIDM2c} (\tau') }{d \tau'} \\  
R_{\rm max}(t)    &=&  R_{\rm max, CDM}(t_f) + \int_{t_f}^{t} d t'  \frac{d R_{\rm max,CDM}(t')}{d t'} +  \int_{t_f}^{t} \frac{dt'}{t_{\rm c}(t')} \frac{d R_{\rm max, SIDM2c} (\tau')}{d \tau'}, 
\end{eqnarray}
\end{widetext}
where $t_f$ is the halo formation time, chosen right after the halo's exponential accretion phase. The derivatives $dV_{\rm max, SIDM2c}(\tau)/d\tau$ and $dR_{\rm max, SIDM2c}(\tau)/d\tau$ are obtained by differentiating the corresponding functions in Eq.~(\ref{eqVR}), with the initial values $V_{\rm max,0}$ and $R_{\rm max,0}$ replaced by $V_{\rm max,CDM}(t')$ and $R_{\rm max,CDM}(t')$, respectively.
After obtaining $V_{\rm max}(t)$ and $R_{\rm max}(t)$ for the corresponding $\tau$, the full \texttt{SIDM2c} density profile can be constructed by solving for the NFW parameters in Eq.~(\ref{eq:tracks}). Further technical details are available in Ref.~\cite{Yang:2024uqb}, which validates this approach using a population of halos from cosmological simulations. This method has also been applied to the \texttt{SIDM Concerto} simulation suite for tests across multiple mass scales~\cite{Nadler:2025jwh}.

\subsection{Incorporating the influence of baryons}

As baryons radiate their energy away from their host halos, they are generically more compact. 
The presence of a deeper baryon potential contracts dark matter and can boost the halo's dynamical evolution. 

To incorporate the effect of baryons on the halo density profile, we adopt a simplified approach by calibrating to the two simulated halos with baryons. 
Both simulations have halo mass $10^{11}~\rm M_{\odot}$ and the stellar content is set by median stellar-to-halo mass and size-mass relations~\cite{2013mnras.428.3121m,2019MNRAS.485..382C}.

Our approach is parametric, utilizing the form factor formula from Ref.~\cite{Yang:2024tba}, defined as
\begin{eqnarray}
{\cal F} &=& \left( \frac{1}{\hat{r}_{\rm eff}} + \frac{ 20\hat{\rho}_H \hat{r}_H^3}{\hat{r}_{\rm eff}(\hat{r}_{\rm eff}+\hat{r}_H)^2} \right)^{-1} \left( 1+1.6\frac{\hat{\rho}_H \hat{r}_H^2}{2} \right)^{-\frac{1}{2}}, \nonumber
\end{eqnarray}
where $\rho_{\rm s}$, and $r_{\rm s}$ are the NFW scale parameters and $\rho_{\rm H}$, and $r_{\rm H}$ are the Hernquist scale parameters. 
The dimensionless quantities are defined as $\hat{\rho}_H=\rho_{\rm H}/\rho_{\rm s}$, $\hat{r}_H=r_{\rm H}/r_{\rm s}$, and $\hat{r}_{\rm eff}=r_{\rm eff}/r_{\rm s}= (1+1.6 \hat{\rho}_H \hat{r}_H^3/2)/(1+1.6 \hat{\rho}_H \hat{r}_H^2/2)$. 
To obtain predictions that incorporate baryonic effects, we simply multiply this factor by the core size $r_c(\tau)$ in the parametric model for DM-only halos.
While schematic, this method accounts for the baryon-induced contraction effect, which scales with the stellar-to-halo mass ratio, and smoothly reduces to the no-baryon limit as $\tau$ approaches zero.  

In the main text, we have illustrate the simulation results for a median concentration halo in Fig.~\ref{fig:dwarfs}a, where baryons have growing cores along with the dark matter. The parametric model predictions are tested against the simulation results at different times. 
Here we provide another example for a more compact halo. 
In Fig.~\ref{fig:rhoctc}c, we compare the parametric model predictions with the simulation results for a $M^{11}~\rm M_{\odot}$ halo ($+1\sigma$ concentration). The profiles, shown in progressively darker colors, correspond to $t=0,3,6,$ and $9$ Gyr, respectively.
While small deviations in shape are present, the model effectively captures the growth of inner densities in the baryonic case (solid lines) compared to the DM-only cases (dashed lines). The baryon profiles themselves (dotted lines) remain largely unchanged in this benchmark.

\section{B. Calculation of strong lensing effects}

\begin{figure*}[htbp]
  \centering

  \begin{overpic}[width=5.5cm]{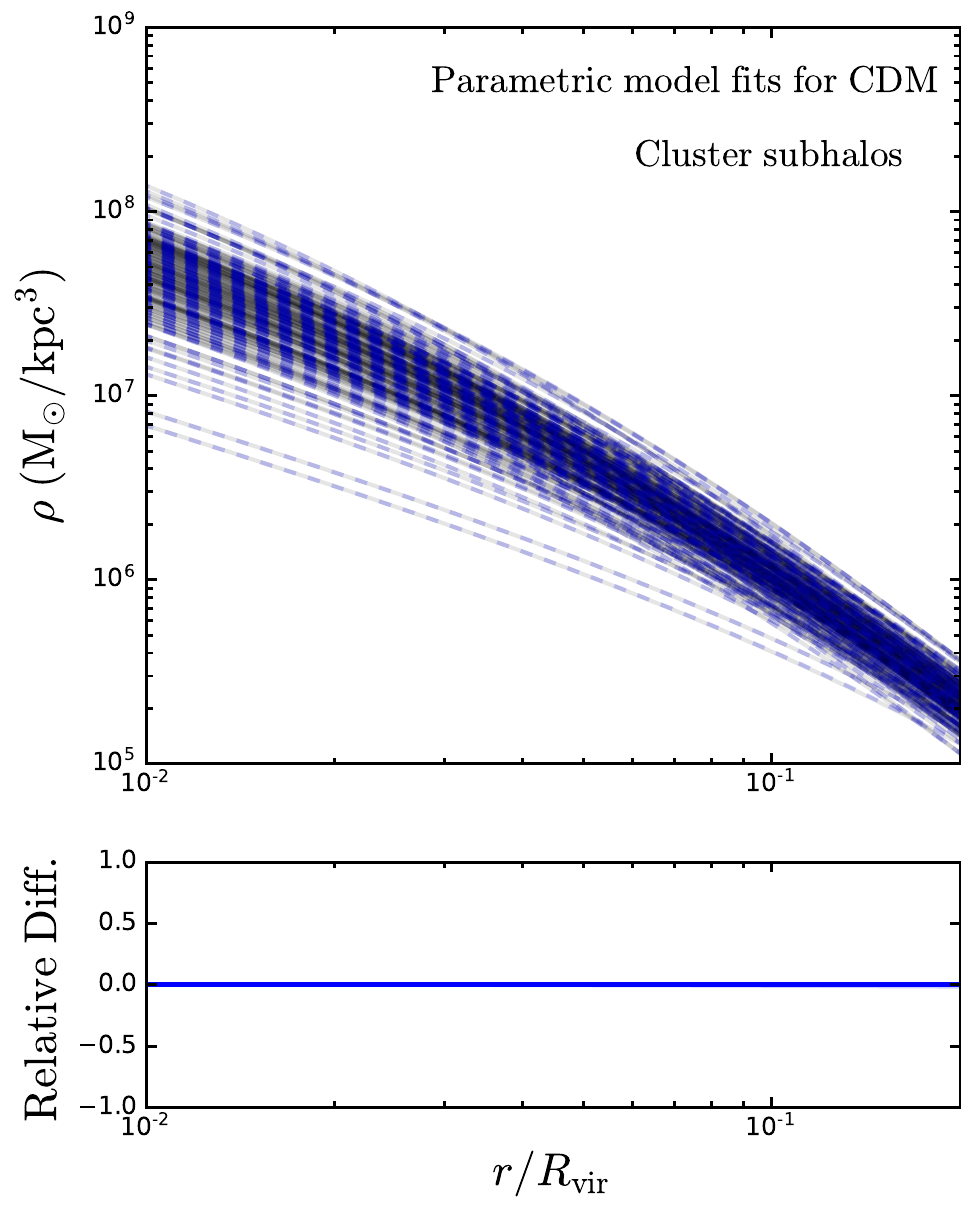}
    \put(2,92){\textsf{\textbf{(a)}}}
  \end{overpic}
  \begin{overpic}[width=5.5cm]{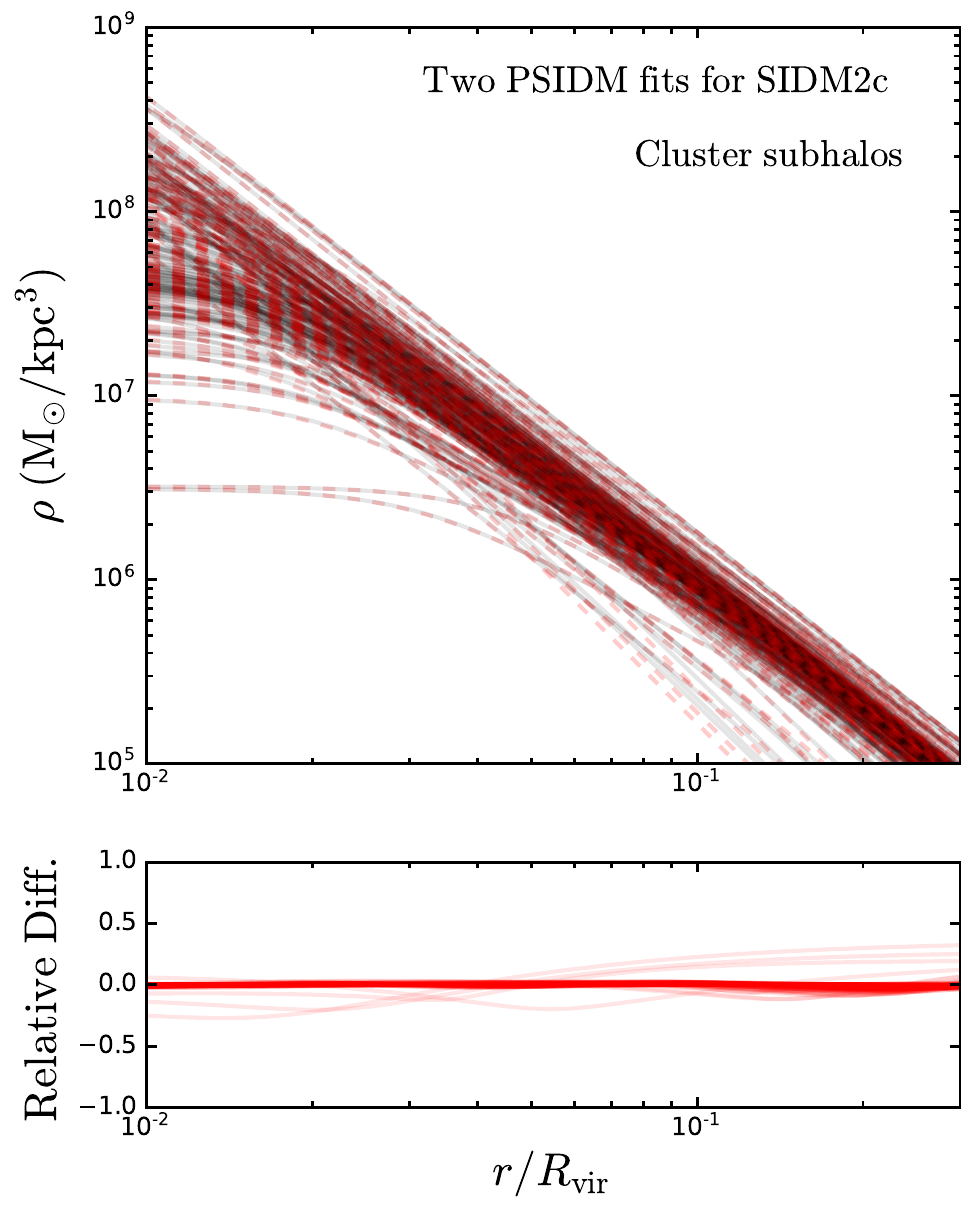}
    \put(2,92){\textsf{\textbf{(b)}}}
  \end{overpic}
  \begin{overpic}[width=5.5cm]{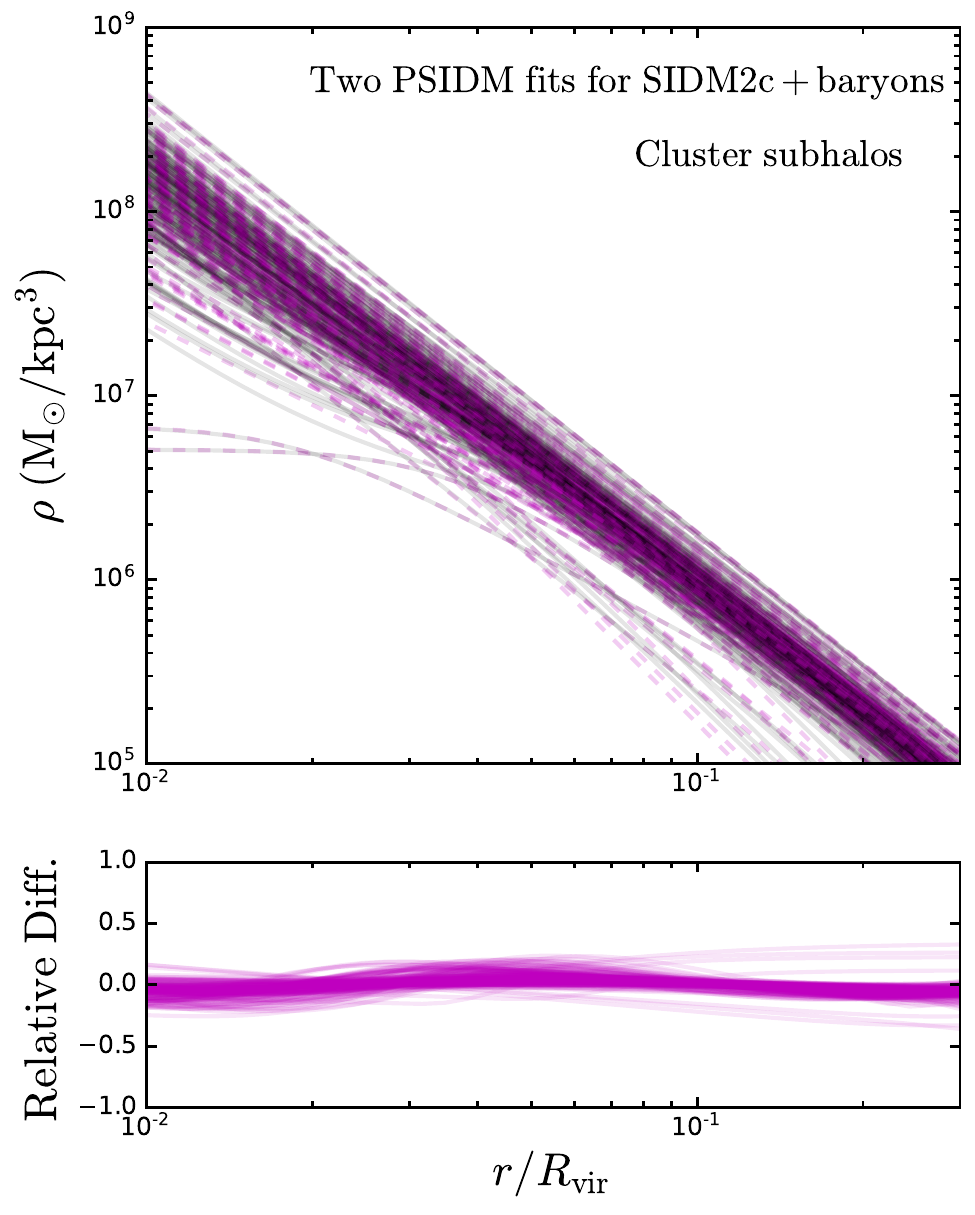}
    \put(2,92){\textsf{\textbf{(c)}}}
  \end{overpic}

  \caption{\label{fig:fits} 
Density profiles of cluster subhalos with $M > 10^{11}\,M_\odot/h$ in \texttt{CDM} (a), \texttt{SIDM2c} (b), and \texttt{SIDM2c+baryons} (c) from parametric model predictions (dashed). Solid curves show corresponding fits using the PSIDM model of Refs.~\cite{Hou:2025gmv}. Subpanels display the relative deviation between simulation and fit, defined as $2\,(\mathrm{Fit}-\mathrm{Sim})/(\mathrm{Fit}+\mathrm{Sim})$. The parametric model accurately captures the density structure of simulated subhalos across a broad mass range.
}
\end{figure*}

After obtaining density profiles from either the \texttt{SIDM2c} parametric model or from simulations, we compute the corresponding lensing signatures. To efficiently obtain lensing predictions for a population of halos, we use the parametric lensing model for SIDM halos developed in Ref.~\cite{Hou:2025gmv}, except for the host halo. 
Although originally developed for one-component SIDM halos, this model can be extended to more general density profiles by treating it as a three-parameter fitting template ($\rho_{s,0}$, $r_{s,0}$, and $\tau$). 
Notably, the gravothermal evolution involves halo inner densities that vary from $r^0$ to $r^{-2.5}$, providing considerable modeling flexibility. A superposition of two such profiles is capable of fitting the more complex structures that arise in the inner halo region.

Specifically, the parametric SIDM (PSIDM) density profile introduced in Ref.~\cite{Hou:2025gmv} has the same parametric form as Eq.~(\ref{eq:PSIDM2c}). The evolution of its halo parameters is governed by the functions given in that reference.
For CDM halos, a single PSIDM profile is sufficient for an accurate fit, as it reduces to the NFW profile in the limit $\tau \to 0$. For \texttt{SIDM2c} halos, we use a superposition of two PSIDM profiles to achieve an accurate representation.

Fig.~\ref{fig:fits} shows the fitting performance for cluster subhalos in three models: from left to right, \texttt{CDM} (blue), \texttt{SIDM2c} (red), and \texttt{SIDM2c+baryons} (magenta). 
Only halos with masses above $10^{11}~\rm M_{\odot}$ are shown. 
The lower panels illustrate the relative difference between the simulated (Sim) and fitted (Fit) profiles, computed as $2(\mathrm{Fit}-\mathrm{Sim})/(\mathrm{Fit}+\mathrm{Sim})$. 
Clearly, the two PSIDM model reproduces the simulated halo profiles with high accuracy, even in the presence of baryonic effects.

Based on the fitted profiles, Ref.~\cite{Hou:2025gmv} provides analytic expressions for the surface density, lensing potential, and deflection angles of each halo. 
The linearity of the lensing equations allows us to superimpose their contributions for more comprehensive lensing predictions.

The cluster host halos are modeled differently from the other halos. 
While the internal ellipticity of other halos contributes subdominantly, the mass and ellipticity of the host halo itself significantly affect the overall lensing signal and must be incorporated. We model this effect by applying a coordinate transformation that introduces a fixed ellipticity into the projected mass distribution.

The ellipticity of each host halo is measured by projecting the associated particles onto the lens plane, taken here as the $x$–$y$ plane of the simulation box (in the fiducial case without rotation). We define the axis ratio $q = b/a$ of the projected mass distribution and compute the elliptical radial coordinate as~\cite{1994A&A...284..285K}
\begin{eqnarray}
    s = \sqrt{q^2 (x\cos\phi + y\sin\phi)^2 + (-x\sin\phi + y\cos\phi )^2}, \nonumber
\end{eqnarray}
where $\phi$ is the angle between the $x$-axis and the semi-major axis of the lens. Substituting $s$ for $r$ in the circular lens model yields an elliptical lens, preserving the enclosed mass within isodensity contours.

We apply a fast Fourier transform to the elliptical surface density to compute the deflection angle maps for the host halos. For the \texttt{CDM}, \texttt{SIDM2v}, and \texttt{SIDMx} clusters, the measured axis ratios are $q = 0.39$, $0.42$, and $0.47$, respectively. These large ellipticities enhance the GGSL effects associated with the host halos.

\section{C. Cluster density profiles and constraint}

\begin{figure*}[htbp]
  \centering

  \begin{overpic}[width=7.2cm]{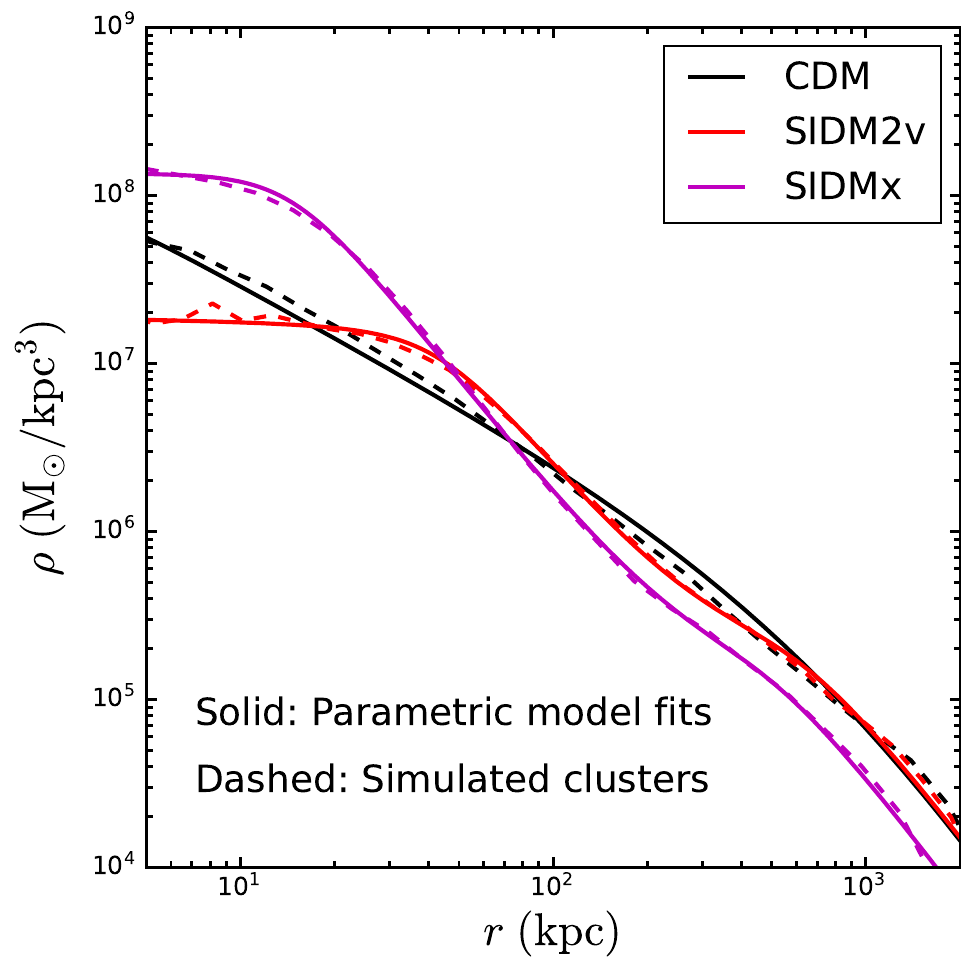}
    \put(2,92){\textsf{\textbf{(a)}}}
  \end{overpic}
  \begin{overpic}[width=7.2cm]{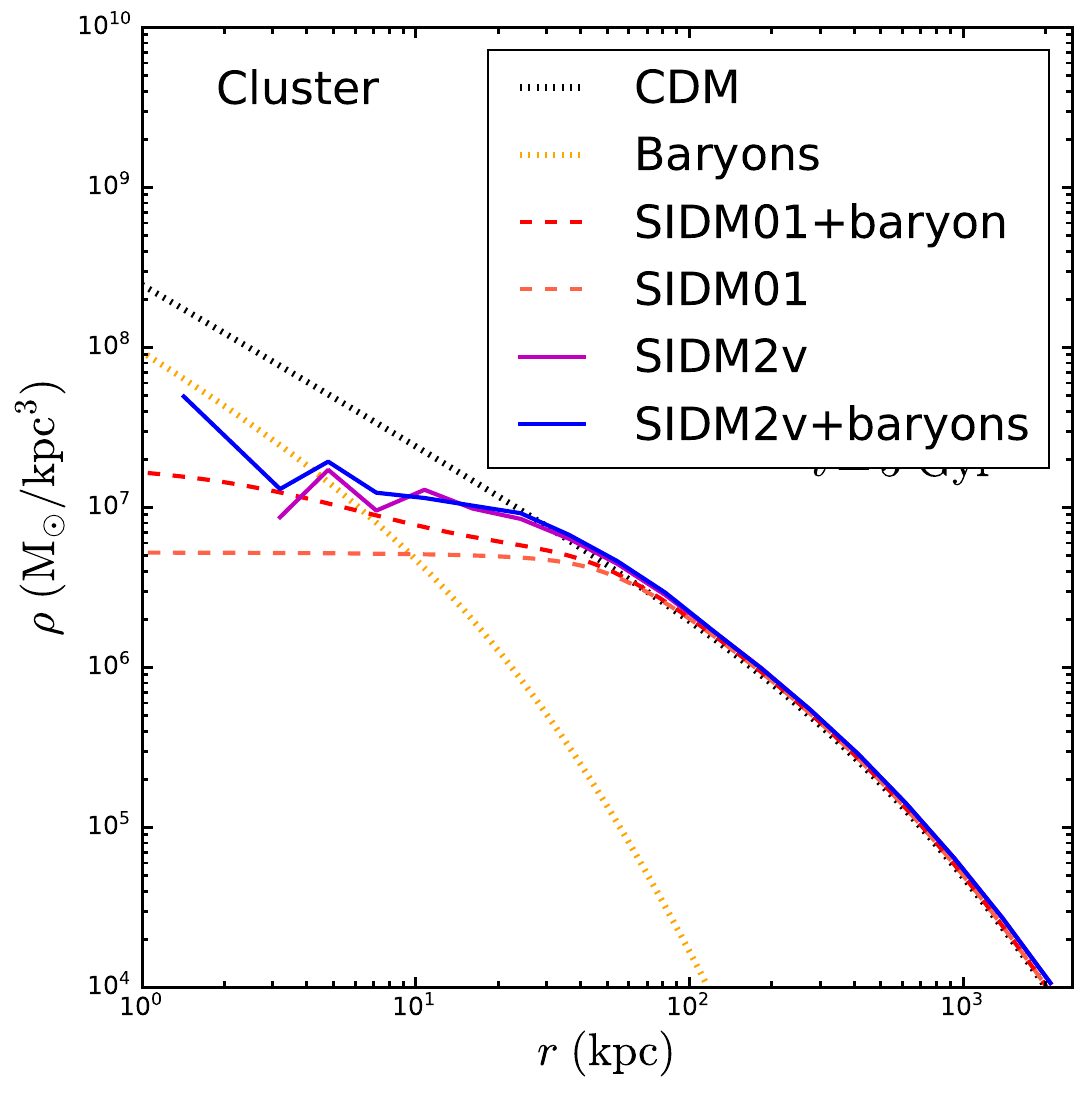}
    \put(2,92){\textsf{\textbf{(b)}}}
  \end{overpic}

  \caption{\label{fig:3clusters} 
\textbf{(a):} Density profiles of cluster host halos in the \texttt{CDM}, \texttt{SIDM2v}, and \texttt{SIDMx} simulations (dashed). The profiles are fitted by two PSIDM density profiles (solid)~\cite{Hou:2025gmv}.  
\textbf{(b):} Comparison of cluster density profiles at $t = 5~\rm Gyr$ for the two-component \texttt{SIDM2v} model (solid) and the one-component \texttt{SIDM01} model (dashed). The initial CDM halo (black) and baryons (orange) are also shown. The inner densities for \texttt{SIDM01} are consistently shallower than those for \texttt{SIDM2v}, both with and without baryons, indicating that \texttt{SIDM2v} remains consistent with cluster-scale constraints.
}
\end{figure*}

The reconstruction of cluster mass profiles has presented one of the most robust constraints on SIDM in clusters, which varies but is generically around $0.1~\rm cm^2/g$~\cite{Andrade:2020lqq,Kaplinghat:2015aga,2019MNRAS.487.1905A}. Core formation in clusters is efficient and only moderately influenced by baryon-induced contraction. For a constant, small self-interaction cross section per unit mass, the rate of gravothermal evolution, quantified by the inverse of the core-collapse time, is proportional to $\rho_{\rm s}^{3/2}r_{\rm s}$, implying that more massive halos evolve more rapidly. Moreover, the stellar mass in clusters generally constitutes only about $0.05\%$ of the total halo mass, and thus contributes negligibly to contraction. 

The presence of mass segregation introduces a mechanism that enhances inner densities, counteracting core formation and thereby alleviating the constraint. In \texttt{SIDM2v}, for example, the inter-species interaction, represented by the magenta curve in Fig.~\ref{fig:xs} around $V_{\rm max} \approx 1600~\rm km/s$, exceeds $0.1~\rm cm^2/g$ and is significantly stronger than the intra-species interaction. This allows it to compete with the efficient core formation.

Fig.~\ref{fig:3clusters}a presents the density profiles of host halos from cosmological simulations. As a result of the merging environment and its interplay with inter- and intra-species scattering, we observe intriguing twists in the density profiles of both \texttt{SIDM2v} and \texttt{SIDMx}, which complicates the assessment of cluster-scale constraints.

To obtain a robust assessment, we perform isolated simulations for the cluster halo under \texttt{SIDM2v}, using the scale parameters of the \texttt{CDM} cluster halo as a reference. 
We consider two contrasting scenarios with and without baryons. The baryonic component, with a mass of $7.47\times 10^{11}~\rm M_{\odot}$, is modeled as a Hernquist profile with $r_{\rm H} = 34.1~\rm kpc$.
These parameters align with the median stellar-to-halo mass relation and median size-mass relation~\cite{2013mnras.428.3121m,2019MNRAS.485..382C}. 
We use $5\times 10^6$ dark matter simulation particles in both cases to better resolve the inner halo regions. Baryons are treated as collisionless stars with particle masses equal to those of the dark matter.

Fig.~\ref{fig:3clusters}b shows the simulation results (solid) at $t = 5~\rm Gyr$, compared with the one-component density profiles (dashed) from the parametric models of Ref.~\cite{Yang:2024tba} with bayrons and Ref.~\cite{yang:2023jwn} without baryons.
The resulting density profiles in \texttt{SIDM2v} are denser than those of the one-component \texttt{SIDM01} model ($\sigma/m = 0.1~\rm cm^2/g$), even in the presence of baryons. 
Since observational constraints depend on the inner mass distribution, our results imply that clusters in \texttt{SIDM2v} are less constrained than those in \texttt{SIDM01}.
In both cases, baryons induce a small contraction that does not alter this conclusion.

\section{D. Uncertainties in the GGSL cross section}

Our simulations and analyses are subject to several systematic and statistical uncertainties. 
In this section, we quantify the dominant uncertainty by analyzing the lens system from different viewing angles, discuss limitations of our numerical methods, and demonstrate the robustness of our primary findings.

To isolate the signal of the relative enhancement from SIDM with mass segregation compared to CDM, we eliminate the inherent randomness in hierarchical structure formation by using identical initial conditions for all simulations. This approach ensures a one-to-one mapping for massive halos across the CDM and SIDM runs, provided SIDM does not substantially alter their accretion histories. 
Therefore, the relative enhancement can be quantified on a halo-by-halo basis. This design intentionally removes halo-to-halo variance, which would otherwise obscure observable signatures. Details such as variations between different cluster halos or the specific spatial distribution of subhalos are rendered marginal in our analysis.

The dominant uncertainty in our GGSL cross section computation arises from analyzing the lens system from different lines of sight. This is significant because strong lensing is sensitive to shear, and the projected ellipticity of our lens changes considerably with viewing angle. In our fiducial setup, the lens lies in the $x$–$y$ plane and is observed along the $z$-axis. We then rotate the viewing angle about the $x$- and $y$-axes in increments of $18^{\circ}$, up to $90^{\circ}$, generating 11 distinct realizations for each cosmological simulation. The statistics of the lensing effects are derived from the ensemble of these realizations. We summarize these results in Table~\ref{tabA1} of the main text. 

We further quantify the statistical differences in GGSL cross sections among the models. This is done using Welch's unequal variances $t$-test, which is suitable for comparing the 11 independent realizations per model that may have different variances. 
The $p$-values and corresponding Gaussian significance in unit of $\sigma$ from this test are presented in Table~\ref{tab:compare}.

The comparison between CDM and Model-2v yields a $2.14\sigma$ difference. All other model comparisons with the CDM baseline show highly significant differences, exceeding $4\sigma$. These results provide strong evidence that our proposed mechanism---SIDM with mass segregation---can produce a significant excess of GGSL events precisely in the regime of observational interest.

\begin{table}[h]
\centering
\caption{Statistical significance of GGSL cross section comparisons. }
\label{tab:compare}
\begin{tabular}{lcc}
\hline
Comparison & $p$-value & significance ($\sigma$) \\
\hline
CDM vs Model-2v & 0.033 & 2.14 \\
CDM vs Model-x & $3.78\times 10^{-6}$ & 4.62 \\
CDM vs Model-2v+baryons & $1.72\times 10^{-5}$ & 4.30 \\
CDM vs Model-x+baryons & $1.03\times 10^{-5}$ & 4.41 \\
\hline
\end{tabular}
\end{table}

\begin{figure}[htbp]
  \centering
  \includegraphics[width=7.2cm]{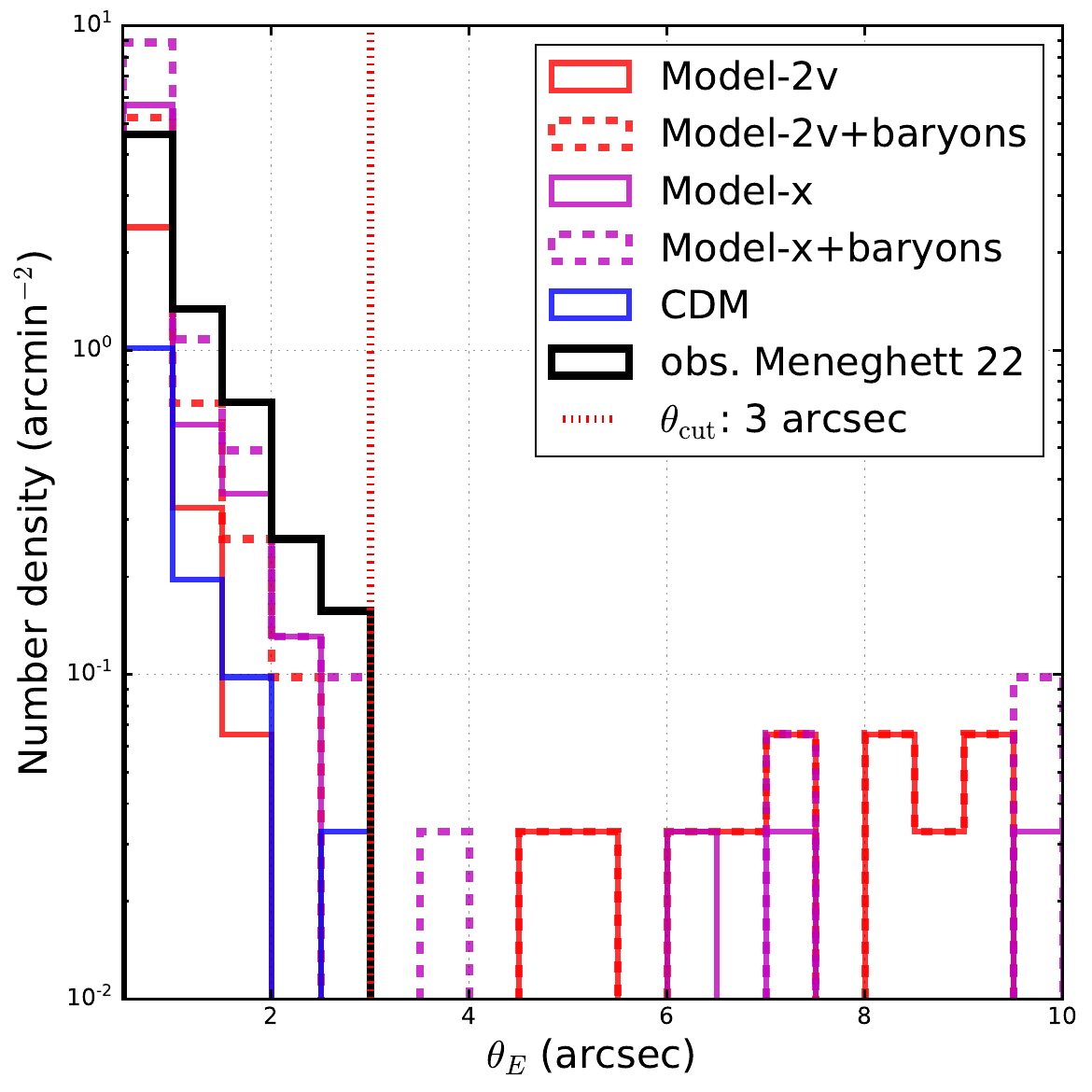}
  \caption{\label{fig:thetaE} Histograms for the Einstein radii of the contributing secondary critical lines in CDM (blue), Model-2v (red), and Model-x (magenta) models. The corresponding cases that incorporate baryonic effects are shown as dashed lines. A vertical dotted line indicates the veto cut $\theta_{\rm cut}=3''$, which is applied to isolate GGSL effects in regions containing observed critical lines. The solid black line shows the observed number density, as taken from Fig.~7 of Ref.~\cite{Meneghetti:2022apr}.
This profile-level comparison demonstrates that SIDM with mass segregation contributes candidates in the range required to explain the GGSL observations. 
}
\end{figure}

\begin{figure*}[htbp]
  \centering
  \begin{overpic}[width=5.5cm]{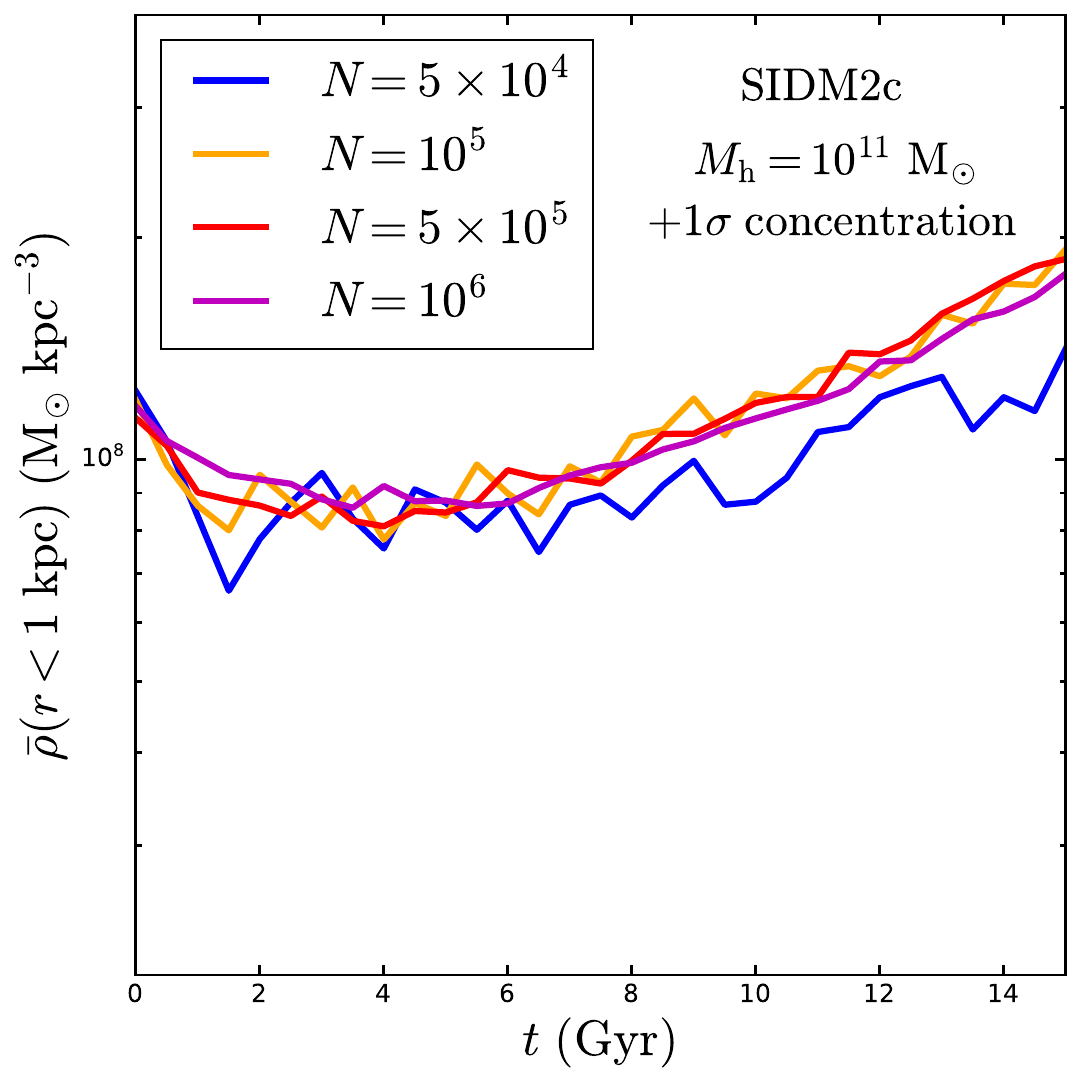}
    \put(1,92){\textsf{\textbf{(a)}}}
  \end{overpic}
  \begin{overpic}[width=5.5cm]{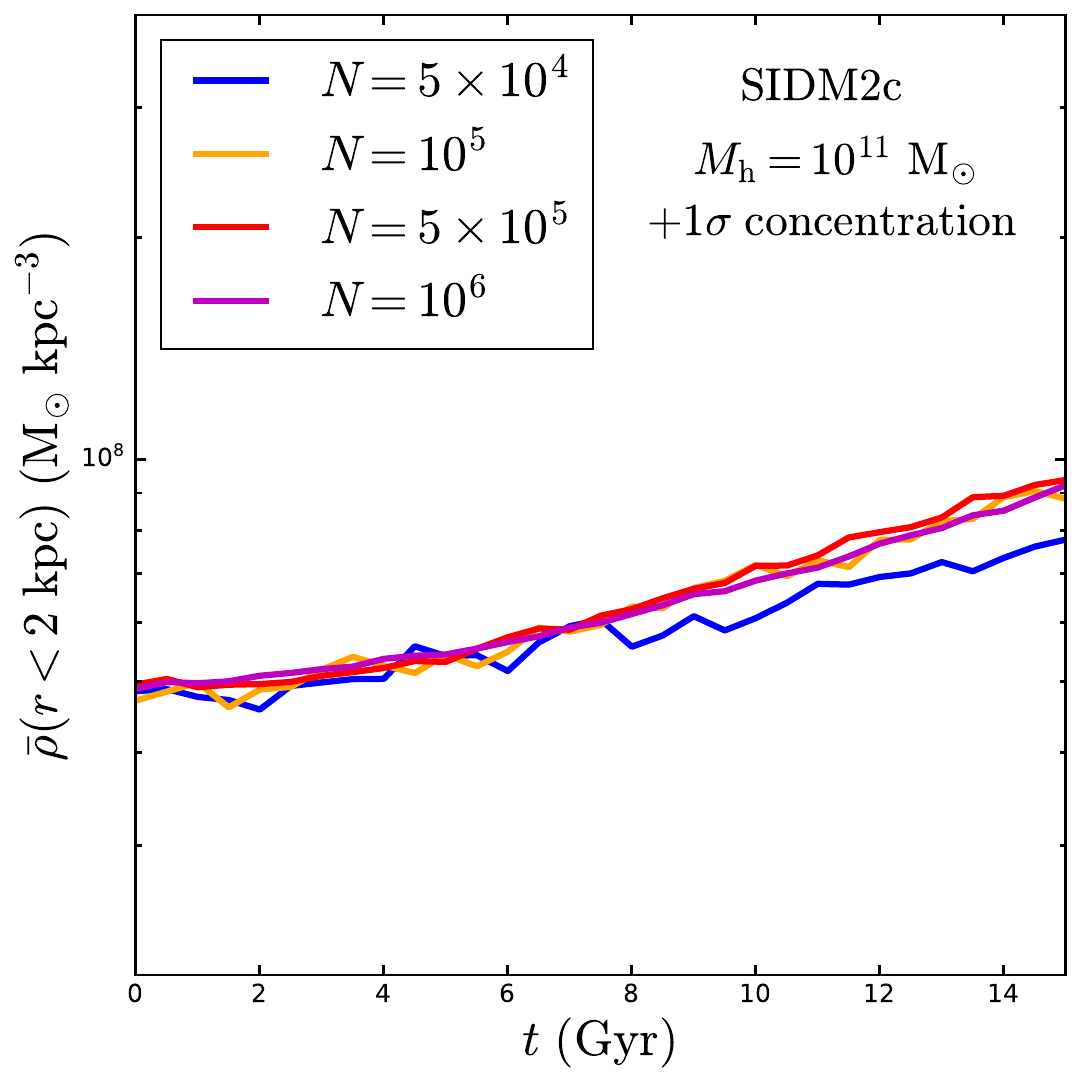}
    \put(1,92){\textsf{\textbf{(b)}}}
  \end{overpic}
  \begin{overpic}[width=5.5cm]{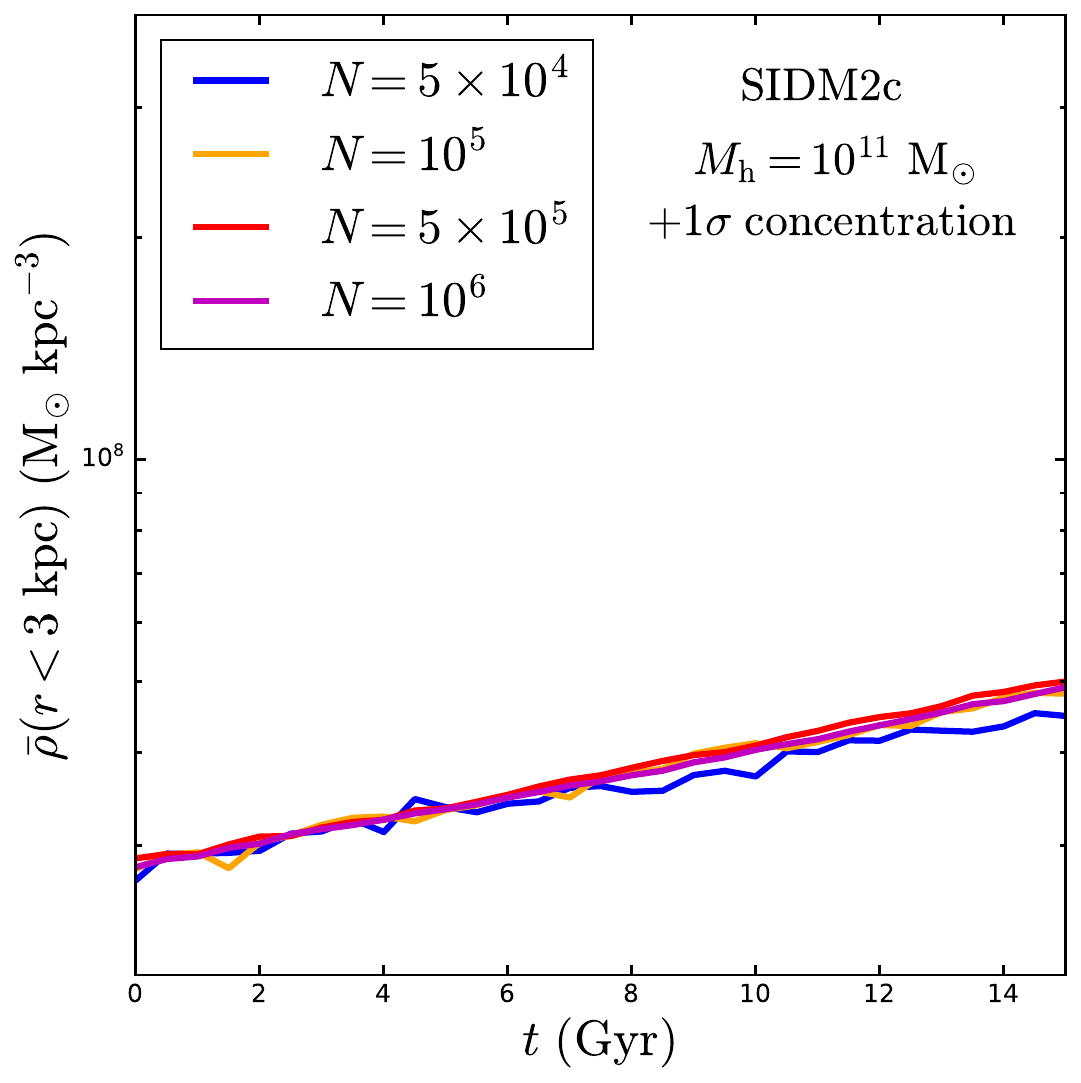}
    \put(1,92){\textsf{\textbf{(c)}}}
  \end{overpic}

  \vspace{2mm}

  \begin{overpic}[width=5.5cm]{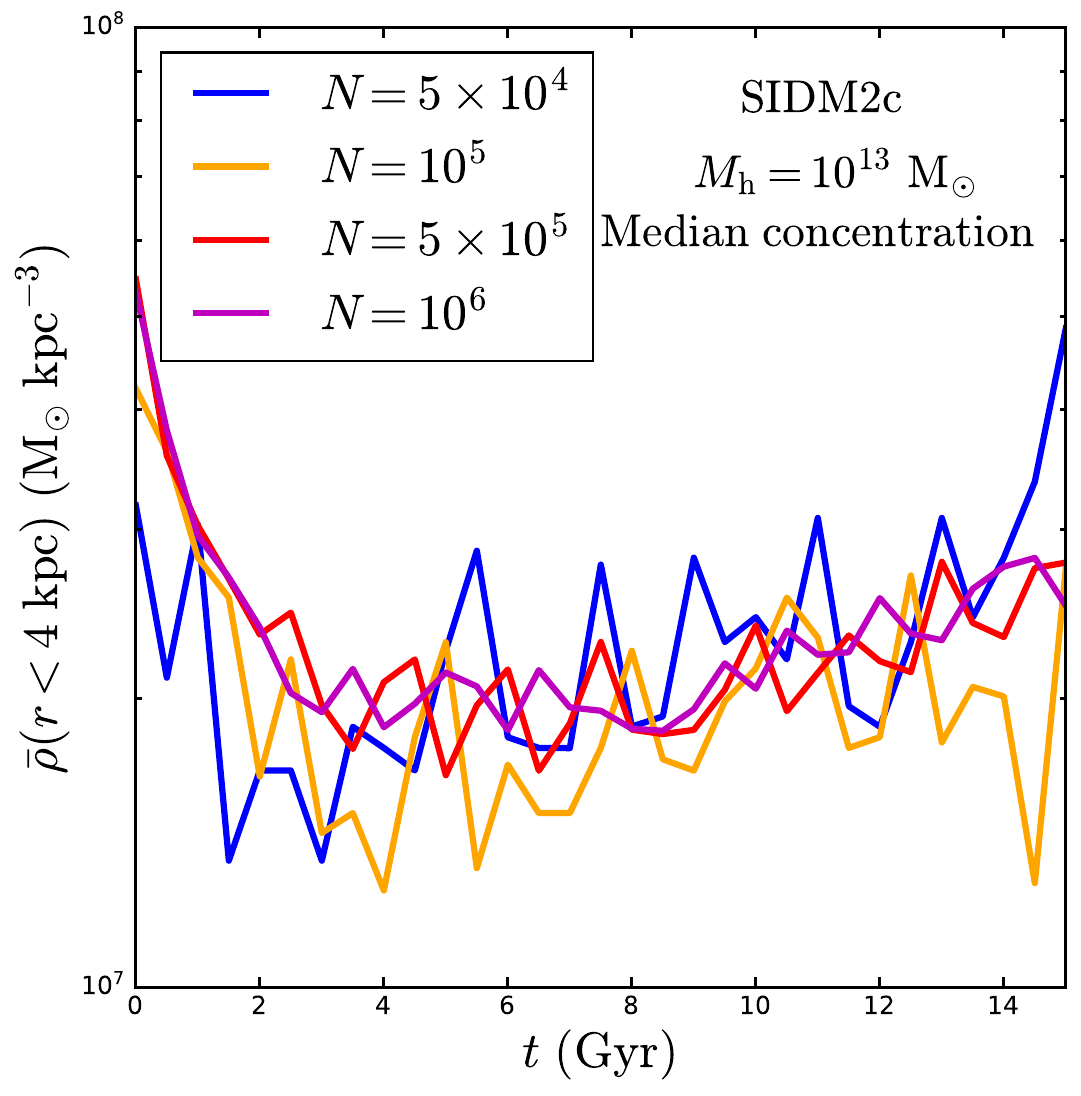}
    \put(1,92){\textsf{\textbf{(d)}}}
  \end{overpic}
  \begin{overpic}[width=5.5cm]{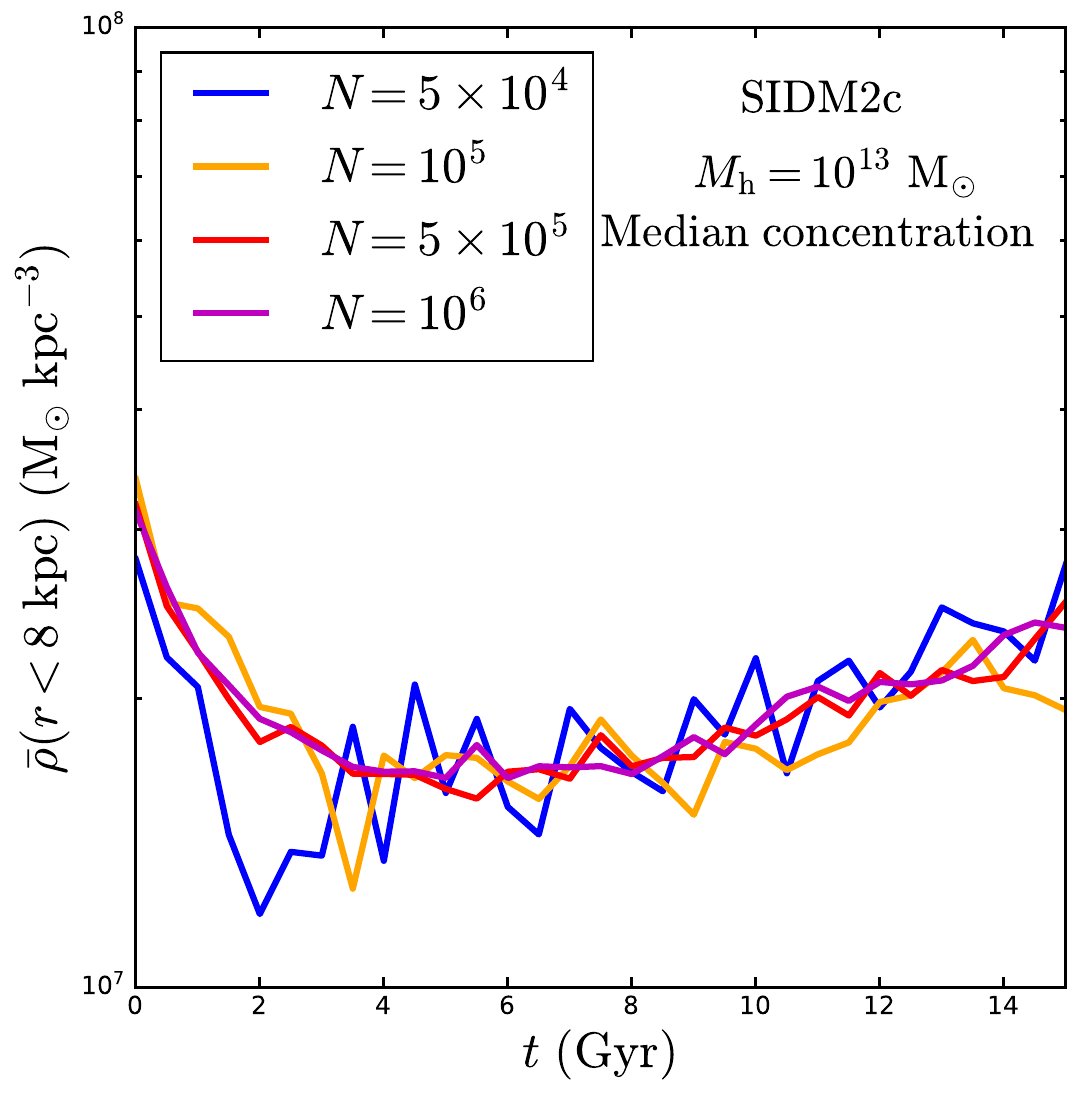}
    \put(1,92){\textsf{\textbf{(e)}}}
  \end{overpic}
  \begin{overpic}[width=5.5cm]{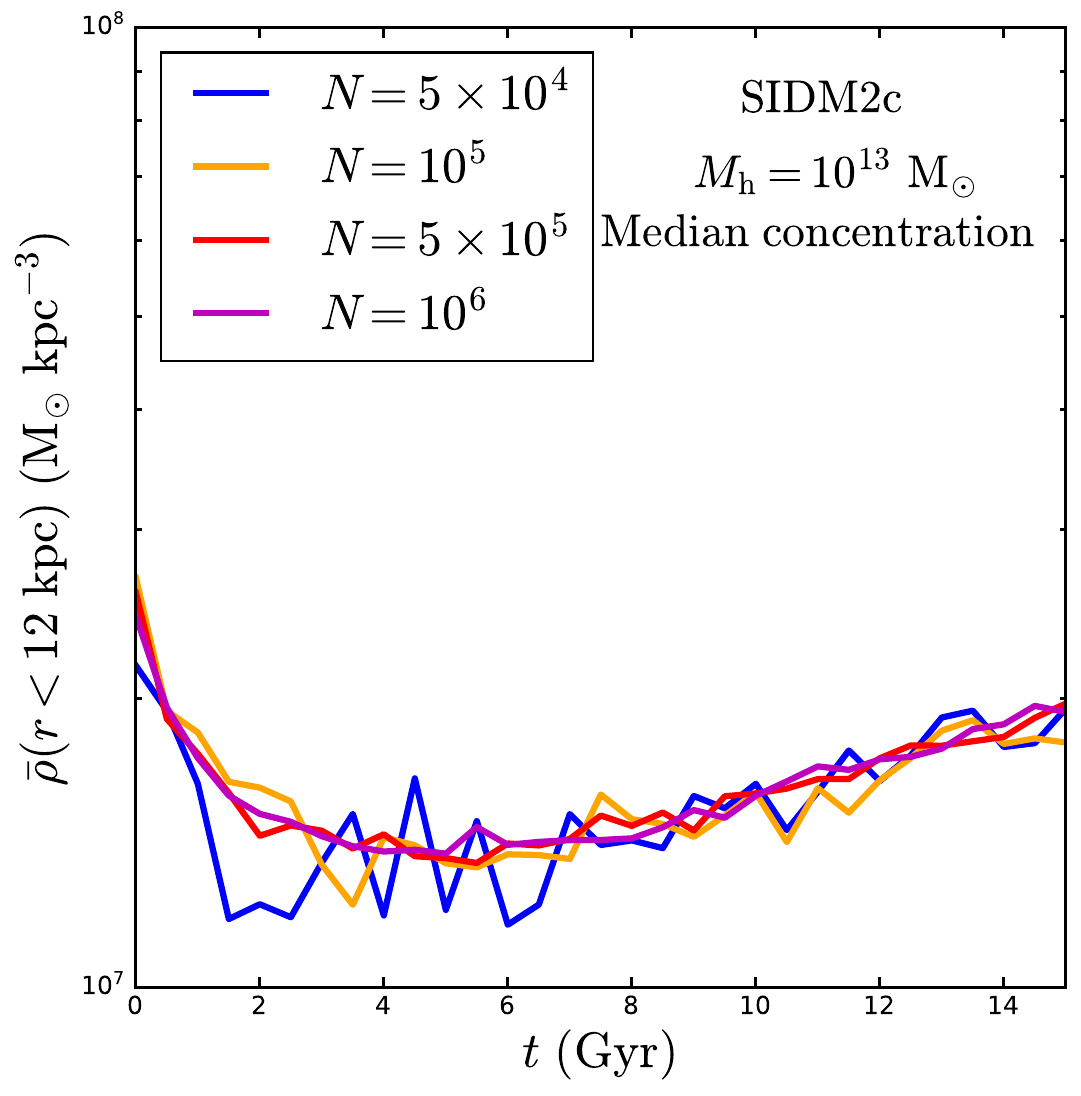}
    \put(1,92){\textsf{\textbf{(f)}}}
  \end{overpic}
  \caption{\label{fig:conv} Convergence test for a $10^{11}~\rm M_{\odot}$ halo with $+1\sigma$ concentration (top) and a $10^{13}~\rm M_{\odot}$ halo with median concentration (bottom). For each halo, we consider $N=5\times 10^4$ (blue), $10^5$ (orange), $5\times 10^5$ (red), and $10^6$ (magenta) simulation particles. The figure shows the evolution of their inner densities at several enclosed radii.
}
\end{figure*}

\section{E. Candidates contributing to the GGSL cross sections}

This section provides a profile-level check that complements the total cross-section comparison presented in the main text and the previous appendix. 
Refs.~\cite{Meneghetti:2022apr,Meneghetti:2023fug,Dutra:2024qac} demonstrate that the subhalos contributing to GGSL predominantly have bound masses in the range $10^{10} - 10^{12}~\rm M_{\odot}/h$ and effective Einstein radii $\theta_E$ between $0.5''$ and $3''$. Here, the Einstein radius is defined as $\theta_E = \sqrt{A_E/\pi}$, where $A_E$ is the area enclosed by the critical curve.

To confirm that the small-scale lenses in our study are consistent with these findings, we identify subhalos with $\theta_E > 0.5''$ and plot their $\theta_E$ distribution in Fig.~\ref{fig:thetaE} for the models in Table~\ref{tabA1}. The number density is obtained by dividing the total number of contributing critical lines by the number of considered systems (11) and by the fiducial calculation area of $100'' \times 100''$. While a few systems have large Einstein radii ($\theta_E > 3''$), the majority of candidates lie in the range $\theta_E < 3''$, in good agreement with observations. Notably, the number densities in our benchmark models, especially those including baryons, show distributions that match the observational data.
This agreement further supports the interpretation that SIDM models with mass segregation can produce subhalo populations consistent with GGSL observations.

The candidates with $\theta_E > 3''$ correspond to massive systems, typically with masses exceeding $10^{13}~\rm M_{\odot}$. A stronger core formation driven by intra-species interaction or AGN feedback could produce shallower inner density profiles, shrinking their Einstein radii. 
Since the physics governing these higher-mass systems operates at a different scale, their properties can be adjusted naturally through either dark matter or baryonic physics. We defer a fully consistent treatment of these systems to future work.

\section{F. Convergence test for halos in the cosmological simulation}

This work involves two types of simulations. The controlled simulations are high-resolution, typically using $10^6$ or more particles per halo. The cosmological simulations, on the other hand, have a fixed average particle mass of $m_0 \equiv (m_{\rm H} + m_{\rm L})/2 = 10^8~\rm M_{\odot}$. This resolution implies that a $5\times 10^{12}~\rm M_{\odot}$ halo is resolved with only $5\times 10^4$ particles.

To demonstrate that our cosmological simulation program can adequately resolve halos down to this mass scale ($5\times 10^{12}~\rm M_{\odot}$), we performed a series of controlled convergence tests. We simulated a $10^{11}~\rm M_{\odot}$ halo (with $+1\sigma$ concentration) and a $10^{13}~\rm M_{\odot}$ halo (with median concentration) at different resolutions: $N = 5\times 10^4$, $10^5$, $5\times 10^5$, and $10^6$ particles.

Fig.~\ref{fig:conv} presents the results. It shows the mean inner density evolution at $1$, $2$, and $3$ kpc for the $10^{11}~\rm M_{\odot}$ halo, and at $4$, $8$, and $12$ kpc for the $10^{13}~\rm M_{\odot}$ halo. The tracks for all resolutions converge with the highest-resolution ($N=10^6$) result. While simulations with fewer particles show increased fluctuations, their overall evolution trend aligns with the high-resolution result.

\begin{table*}[htbp]
\begin{center}
\caption{Parameters of the controlled N-body simulation halos.}
\label{tab:consim}
\begin{tabular}{cccccccc}
\hline
\hline
Simulation  & $M_{\rm h} ({\rm M_{\odot}})$ & $\rho_{\rm s} ({\rm M_{\odot}/kpc^3})$ & $r_{\rm s} ({\rm kpc})$ & $M_s ({\rm M_{\odot}})$   &  $r_{\rm H} ({\rm kpc})$  & SIDM model  & $N_{\rm part}$ \\
\hline
M02         & $2\times 10^{10}$ & $1.03\times 10^7$ &  $4.53$       & 0                       &  0         & \texttt{SIDM2c} &    $10^6$ \\ \hline
M02p1       & $2\times 10^{10}$ & $1.94\times 10^7$ &  $3.51$       & 0                       &  0         & \texttt{SIDM2c} &    $10^6$ \\ \hline
M02p2       & $2\times 10^{10}$ & $3.71\times 10^7$ &  $2.73$       & 0                       &  0         & \texttt{SIDM2c} &    $10^6$ \\ \hline
M05         & $5\times 10^{10}$ & $4.05\times 10^7$ &  $3.58$       & 0                       &  0         & \texttt{SIDM2c} &    $10^6$ \\ \hline
M08         & $8\times 10^{10}$ & $3.58\times 10^7$ &  $4.39$       & 0                       &  0         & \texttt{SIDM2c} &    $10^6$ \\ \hline
M11         & $10^{11}$     &  $6.89\times 10^6$ &  $9.10$       & 0                     &  0         & \texttt{SIDM2c} &    $10^6$ \\ \hline
M11         & $10^{11}$     &  $6.89\times 10^6$ &  $9.10$       & 0                     &  0         & \texttt{SIDM03} &    $10^6$ \\ \hline
M11p1       & $10^{11}$     &  $1.29\times 10^7$ &  $7.07$       & 0                     &  0         & \texttt{SIDM2c} &    $10^6$ \\
\hline
M11m1       & $10^{11}$     &  $3.73\times 10^6$ &  $11.7$       & 0                      &  0         & \texttt{SIDM2c} &    $10^6$ \\ \hline
M11B         & $10^{11}$     &  $6.89\times 10^6$ &  $9.10$       & $10^9$                &  0.768         & \texttt{SIDM2c} &    $10^6+8281$ \\ \hline
M11p1B       & $10^{11}$     &  $1.29\times 10^7$ &  $7.07$       & $10^9$                &  0.768         & \texttt{SIDM2c} &    $10^6+8420$ \\ \hline
M15       & $1.79\times 10^{15}$     &  $3.09\times 10^5$ &  $860$       & 0                &  0         & \texttt{SIDM2v} &    $5\times 10^6$ \\ \hline
M15B      & $1.79\times 10^{15}$     &  $3.09\times 10^5$ &  $860$       & $7.47\times 10^{11}$                &  $34.1$         & \texttt{SIDM2v} &    $5\times 10^6 + 1556$ \\ \hline
\end{tabular}
  \vspace{0.5em}
  \parbox{\linewidth}{\footnotesize
The labels ``p1'', ``p2'', and ``m1'' in the benchmark names refer to simulations with concentrations of $+1\sigma$, $+2\sigma$, and $-1\sigma$ relative to the median. The labels ``B'' imply the presence of baryons. 
Halo masses are determined as the mass enclosed within radii where the density reaches 200 times the critical density ($R_{200}$), based on the scale density ($\rho_{\rm s}$) and scale radius ($r_{\rm s}$) of the NFW profile. The total halo masses in the simulations are slightly higher due to the smooth truncation of the density profiles. Stellar masses are calculated as total masses using the scale density ($\rho_{\rm H}$) and scale radius ($r_{\rm H}$) of the Hernquist profile. Dark matter and stellar simulation particles have equal mass. $N_{\rm part}$ denotes a summation of their numbers. 
}
\end{center}
\end{table*}

\section{G. Summary of controlled simulations}

We summarize in Table~\ref{tab:consim} the parameters of the high-resolution controlled simulations in this work.

\end{document}